\documentstyle[prl,aps,epsf,preprint]{revtex}

\newcommand{\tJ}{$t$-$J$\ }
\newcommand{\Tc}{$T_c$\ }

\begin{document}
\draft
\title{Theory of spin excitations in undoped and underdoped cuprates}
\author{Don H. Kim$^{*}$  and  Patrick A. Lee}

\address{Department of Physics, Massachussetts Institute of Technology,
Cambridge, MA, 02139}
\date{\today}

\maketitle
\tighten
\begin{abstract}
We consider the magnetic properties of high \Tc cuprates
from a gauge theory point of view, with emphasis on
the underdoped  regime.  Underdoped cuprates possess certain antiferromagnetic
correlations, as evidenced, for example, by
  different temperature dependence of the Cu and
O site NMR relaxation rates, that are not captured well by
slave boson mean field theories of the \tJ model.  We show that the inclusion
of gauge fluctuations will remedy the deficiencies of the mean field theories.
  As a concrete illustration of the gauge-fluctuation
restoration of the antiferromangetic correlation
and the feasibility of the $1/N$ perturbation theory, the Heisenberg spin
chain is analyzed in terms of
a 1+1D U(1) gauge theory with massless Dirac fermions.
The $1/N$-perturbative
treatment of the same  gauge theory in 2+1D (which can be motivated from
the mean field $\pi$-flux phase of the Heisenberg model) leads to a dynamical
mass generation corresponding to an antiferromagnetic ordering. On the other
hand, it
is argued that in a similar gauge theory with an additional coupling to
a Bose (holon) field, symmetry breaking does not occur, but antiferromagnetic
correlations are enhanced, which is the situation in the underdoped cuprates.
\end{abstract}



\section{Introduction}

The essence of the physics of high $T_c$ cuprates boils down to the problem
of how to treat the
dual nature of the electrons which form local moments in the insulator, and 
yet make up a Fermi surface when doped with $\sim 15\%$ holes.  This
problem was brought into sharp
focus with the discovery of anomalous properties in the physics of the
``underdoped''
region which lies between the antiferromagnetic insulator and the optimally
doped superconductor.
 How does the Fermi surface evolve from small hole pockets
near ${\bf k}=(\pm\pi/2,\pm\pi/2)$ in
a slightly doped antiferromagnet to the full
Fermi surface obeying Luttinger's theorem in the optimally doped materials?
What are the magnetic properties in this intermediate doping region?
 Experimentalists have already answered a substantial part of these questions.
In particular, the angle-resolved photoemission spectroscopy
(ARPES)  has shown the existence in the normal state
 of a gap with the same anisotropy as the
 d-wave gap of the superconducting state\cite{photo}.
  Low-lying excitations are observed
along a patch near $(\pm\pi/2,\pm\pi/2)$
[``Fermi surface segments''], but the Fermi
surface, in the veritable sense of the word,
does not exist. The ARPES results might
have been accepted without much grudge simply as a
plausible  interpolation between the antiferromagnet and
the optimally doped superconductor, had our understanding of metals not been
so entrenched  in the Fermi liquid theory; the notion of a metal without
a Fermi surface is a serious embarrassment.
At the same time,  gaplike suppression of spin excitations are seen in
NMR\cite{taki}:
the Knight shift and spin-lattice relaxation rates all decrease with decreasing
temperature below  certain temperatures.

Gaplike features in the underdoped cuprates might remind us of the
spin liquids --- the liquid of spin singlets.
 Yet the devil is in the details, and the underdoped cuprates deviate
significantly from  ``gapped spin liquids'' like  spin 1/2 ladders and
integer spin chains.  The latter have an even number of spins per unit cell
and the ground state is a spin singlet.  They
are characterized by a
clear gap  to the lowest triplet excitation
 that is inversely related to the correlation length; this gap
can be seen in  inelastic
neutron scattering.  The magnetic
responses like uniform susceptibility and the NMR
relaxation rate as a function of temperature have  activated behaviors.
Many of them can be satisfactorily described in terms of the
 ``quantum disordered''  phase of the
 nonlinear sigma model ${\cal L}=\frac{1}{g} (\partial_{\mu}{\bf n})^2,
g > g_c$\cite{olavchak} which can be also
understood in terms of the $CP^{N-1}$ model involving {\it bosonic}
``spinons'' ($z$ fields)
--- these spin 1/2 excitations are said to be confined, as they do not
appear in the basic
physical spectra\cite{chubukov}.  In other words, a description of the
excitations in
terms of fluctuating spin 1 objects is most natural for them.

On the other hand, in the underdoped cuprates, inelastic neutron scattering
 does not show a ``full gap''\cite{chakshort}.
The decrease of the  Knight shift
 with decreasing temperature looks more like a
power-law.  On the whole, the magnetic excitation spectrum of the
cuprates seems to display a curious mixture of singlet and
antiferromagnetic correlations.  There are  evidences for
antiferromagnetic correlations from the {\bf Q}-space scan of
neutron scattering
cross section, and from the difference in the temperature dependence
of the NMR  relaxation rates  : the Oxygen $1/T_1T$ (which has
little contribution
from spin excitations near wave vector ${\bf Q}=(\pi,\pi)$)
monotonically decreases with decreasing
temperature, while the Copper $1/T_1T$
 (which weighs $(\pi,\pi)$ spin excitation
strongly) increases with decreasing temperature until around 150K and then
falls\cite{taki}.

The antiferromaget-singlet debate has enormous ramifications
 for  theories of high \Tc cuprates (for a succinct review, see
Ref.\onlinecite{nagaosa}).
Some theorists\cite{antipara} have advocated the picture of
Fermi liquid quasiparticles exchanging
``antiparamagnons''  for understanding the anomalous normal state
properties and the superconductive pairing.
 Such a view is not shared here.
Instead of viewing the antiferromagnetic fluctuations  as the cause
of  superconductivity in a BCS-like scenario, in this paper
we would rather regard them  as
 a residual but important
consequence of local
repulsive interactions that lead to superconductivity in the presence of
doped holes, a part of Nature's conspiracy to find
a compromise between a magnetic ground state and an itinerant metallic state.
 This line of thinking goes back to Anderson's seminal 1987
paper\cite{anderson1} on the resonating valence bond (RVB) theory, in
which he reasoned that doped holes may propagate coherently in a liquid
of spin singlets.

Theoretical attempts to realize Anderson's RVB picture are based on
strong coupling models, such as the one band Hubbard model or the
\tJ model\cite{zhangrice}.
 These models are considered to contain some essential physics
of the cuprates at  appropriate parameter values $U/t$ or $J/t$.
The \tJ model, the simpler of the two,  captures in a transparent way what is
believed to be the basic physics, namely the competition between the
magnetic exchange and the delocalization energy of holes.
The no-double-occupancy constraint in the \tJ model can be taken care of
by writing the electron operator as a composite of a neutral
fermion (spinon) and a spinless boson (holon)
[$c_{i\sigma}^{\dagger}=f_{i\sigma}^{\dagger}b_i$] and demanding
each site be occupied by either a fermion or a boson
$(\sum_{\sigma}f_{i\sigma}^{\dagger}f_{i\sigma}+b_i^{\dagger}b_i=1$).
The theory then contains four-particle interactions which can be decoupled
by introducing ``mean fields'' like $\chi_{ij}=
\langle f_{i\sigma}^{\dagger}f_{j\sigma} \rangle $,
 $\Delta_{ij}=\langle f_{i\uparrow}f_{j\downarrow}-
f_{i\downarrow}f_{i\uparrow}\rangle$, and $\eta_{ij}=\langle b_i^{\dagger}b_j
\rangle$.
Within the mean field approach,
Kotliar and Liu,  and Fukuyama and
coworkers\cite{kotliar}  have studied the phase diagram of the \tJ model.
At low doping (and below some temperature scale),
it was found that the phases in which the fermions are
paired into d-wave singlets ($\Delta_{i,i+x}=-\Delta_{i,i+y}
\neq 0$)are favored.
 Depending on whether the bosons
are condensed, they could be superconducting (SC phase) or normal (``d-wave
RVB phase'').

As noted by Rice\cite{rice_spin} and others, the fermionic mean field theory
captures some important features of
the spin gap phenomena in the underdoped cuprates
that  refuse clear-cut
characterization in terms of a well-defined correlation length and
relaxation times.
The theory describes some kind of quantum spin liquid, but
unlike gapped spin liquids, there is a particle-hole
(spinon-antispinon) continuum, which would create some spectral weight
for magnetic excitations at
arbitrarily low energy.  More specifically, the Dirac spectrum
($\epsilon({\bf k})=v|{\bf k}|$) of the
fermionic quasiparticles in the d-wave RVB phase gives
the Knight shift $K\sim T$ and the Oxygen site
NMR
relaxation rate $1/T_1\sim T^3$,
in rough agreement with experiments in the underdoped cuprates.
Moreover,
the absence of the gap in the charge response (for example, the
 in-plane optical
conductivity) could be explained simply, since the spin and charge
degrees of freedom are separated, {\it  i.e.} the spin is carried by fermionic
spinons while the charge is carried by bosonic holons.

A Dirac-type spectrum as in the d-wave RVB phase was also
found by Affleck and Marston who considered  the $\pi$-flux
phase\cite{flux} as
 a possible spin liquid ground state of the cuprates.
It turned out that at half filling
the d-wave phase with $|\Delta_{ij}|=|\chi_{ij}|$
 is equivalent to the flux phase, due to a local SU(2) symmetry\cite{affzou}.
 Wen and one of us reasoned that this symmetry might still
 be a pretty good (and important) symmetry at
small dopings, and came up with a slave boson theory that respects the SU(2)
symmetry even away from half filling by introducing an SU(2)
 doublet of slave bosons,
hoping to get a better description of underdoped cuprates\cite{wenlee}.
In this theory, the mean field corresponding to the ``spin gap'' phase of the
underdoped cuprates was identified as the  ``sFlux phase'' which can be
considered a combination of the d-wave RVB phase and the staggered flux
phase\cite{fczhang}
of the U(1) theory.  This phase also has
fermions with a Dirac spectrum, but {\it in contrast to the d-wave RVB phase
of the U(1) mean field theory}\cite{u1instab},
{\it  the fluctuations around this mean field
include a  massless gauge field which is expected to affect strongly the
magnetic and transport properties of the system}\cite{leeetal}.
  With the inclusion of a
residual attraction
between bosons and fermions, the sFlux phase was shown to reproduce the gross
features of the ARPES, such as the Fermi surface segments near $(\pm
\pi/2,\pm \pi/2)$ and a  large gap at $(\pi,0)$.

Despite the successes,    the mean field treatments (both the
SU(2) theory
and its predessesors) are  unsatisfactory in several respects.  For
example, it is not clear how the spin gap phase is connected to the
N\'{e}el ordered phase at zero doping.  {\it The mean field ansatz loses a lot
of antiferromagnetic correlation}; within the mean field theory,
the Copper site $1/T_1T$ has a similar behavior as the Oxygen site $1/T_1T$,
in disagreement with experiments\cite{taki}.  Attempts to fix the
problem by some kind of RPA cannot  produce, in a natural manner,
  different
temperature
scales which mark the decrease of the Copper and Oxygen site relaxation
rates\cite{fukuyama2}.
  Another serious question is the role of gauge fluctuations around the
mean field solution which had not been studied carefully so far.  In fact, a
strong gauge fluctuation might destroy the mean field picture altogether,
in which case we have to re-identify the  elementary excitations of the
theory.

 In this paper, we look into these questions.  The basic point is that
the gauge fluctuations ignored at the mean field level strongly enhance
antiferromagnetic correlations.  The gauge fluctuation could be so strong
that the elementary excitations of the mean field theory disappear completely
 from
the low energy spectrum.  This is believed to be the case with the
``U(1) $\pi$-flux
phase''\cite{flux}
 description of the undoped cuprates (antiferromagnet), which is a
theory of massless Dirac fermions  coupled to a U(1) gauge
field (massless QED3).  The QED3 can be treated in the
$1/N$
perturbation theory.  For physical $N (=2)$, a dynamic mass generation
and spontaneous symmetry breaking corresponding to N\'{e}el ordering would
occur\cite{literature},
while for large enough $N$, the theory would still describe some kind of
a spin liquid.
 The true low energy excitations
of the symmetry-broken case
are  recognized as the Goldstone bosons--``mesons'' which are a bound state
of a particle and
an antiparticle (spinon \& antispinon).

We note that the local SU(2) symmetry  at half filling can be
utilized to write a theory of the undoped system in terms of
massless Dirac fermions coupled to nonabelian (SU(2)) gauge
fields\cite{affzou,leeetal}.  Since the U(1) theory (at least initially) has
one massless gauge field  while the SU(2) theory has three,
they look quite different.  Nonetheless,
 as long as the gauge fluctuations are treated exactly, the two theories
 should
lead to the same physics, namely the N\'{e}el ordering.

 As regards the {\it underdoped} cuprates, we believe that
the most promising starting point is the sFlux phase of the
SU(2) theory\cite{leeetal}, which has massless Dirac fermions
coupled to a massless
U(1) gauge field as in the Affleck-Marston flux phase\cite{flux},
but also has nonrelativistic bosons(holons) coupled to the
same gauge field.
We shall  argue that, due to this additional coupling, the
gauge fluctuations will not destroy the essential validity of the mean field
picture, though the
picture of antiferromagnetic spin excitations will be modified (improved).
 The coupling to the bosons would result in the screening of the time component
of the gauge field which will prevent N\'{e}el ordering.
The gauge field will nevertheless  mediate an
attraction between spinons and
antispinons  and try to create a bound state with momentum $\sim (\pi,\pi)$,
 but due to the particle-hole continuum,
 this will appear  only as a broad resonance.  This can be
viewed as a Goldstone boson precursor mode that comes down in energy as the
transition is approached (as the boson density is reduced).
 The recent
neutron scattering in underdoped
cuprates which sees a broad peak in ${\bf Q}\approx (\pi,\pi)$
magnetic response
whose energy scale is roughly proportional to doping might be consistent
with this point of view\cite{keimer}.
 We shall also discuss the issue of confinement,
as there are
lingering questions about the fate of ``spinons'' in the case of strong
coupling gauge theories.

In order to illustrate some aspects of the foregoing ideas
 more concretely ({\it in particular the gauge-fluctuation restoration
of antiferromagnetic correlation and the feasibility of a U(1) gauge
theory description of quantum antiferromagnets}),
 we'll first reexamine the well known spin
half chain from the point of view of the  Schwinger model.

\section{Lessons from Spin Chain}
We begin with a discussion of the 1d Heisenberg model as an example
where the idea of mean
field theory plus gauge fluctuation can be applied to an exactly soluble
model.  A number of
authors have noted that what emerges is the Schwinger model with two flavors
\cite{Wiegmann,hosotani,mudry},
and abelian \cite{hosotani} and nonabelian
\cite{mudry} bosonization
methods have been applied to solve for the continuum limit of this model.
Here we solve the model
using a slightly different technique, i.e., chiral rotation, which focusses
on the role of gauge
fluctuations as a way of correcting the underestimate of the
antiferromagnetic correlation in mean
field theory.  We then show that qualitatively similar results are achieved in
$1/N$ perturbation theory
of the gauge fluctuations.  This gives us encouragement that a similar
approach may be reasonable
in the two-dimensional case.

\subsection{RVB theory of  spin one-half chain}
The Heisenberg model ($H=J\sum_{<ij>}{\bf S}_i\cdot {\bf S}_j$)
 in the fermion representation of the spin can be written
\begin{equation}
H=-\frac{J}{2}\sum_{<ij>} f_{j\alpha}^{\dagger}f_{i\alpha}
f_{i\beta}^{\dagger}f_{j\beta}
\end{equation}
with the constraint $\sum_{\alpha}f_{i\alpha}^{\dagger}f_{i\alpha}=1$.
The 4-fermion interactions and the constraint    can be
handled by the introduction of a Hubbard-Stratonovich field and a Lagrange
multiplier, which gives
\begin{equation}
H=-J/2\sum_{<ij>}(\chi_{ij}f_{i\alpha}^{\dagger}f_{j\alpha}+{\rm h.c.})+
i\sum_i\lambda_i (f_{i\alpha}^{\dagger}f_{i\alpha}-1).
\end{equation}
Within the mean field theory, $\chi_{ij}=\chi,\lambda=0$, hence the
mean field hamitonian in the k-space is
\begin{eqnarray}
H_{mf}&=&-\chi J \sum_k \cos(k)f_{\alpha k}^{\dagger}f_{\alpha k} \\
      &=&-\chi J\Sigma_k^{'} \cos(k)(f_{\alpha k}^{\dagger}f_{\alpha k}
-f_{\alpha, k-\pi}^{\dagger}f_{\alpha, k-\pi})\\
&=& -\chi J  \Sigma_k^{'} \cos(k)(f_{\alpha ek}^{\dagger}f_{\alpha ok}
+f_{\alpha ok}^{\dagger}f_{\alpha ek}).
\end{eqnarray}
Here $\Sigma_k^{'}$ denotes  sum over the magnetic BZ, say $0<k<\pi$,
and $f_{ek}, f_{ok}$ are even and odd site operators ($f_{ek}=
\frac{1}{\sqrt{2}}(f_{k}+f_{k-\pi})=\sqrt{\frac{2}{L^2}}\sum_{j\,
{\rm even}}e^{ikj}f_j$,
$f_{ok}=\frac{1}{\sqrt{2}}(f_k-f_{k-\pi})=\sqrt{\frac{2}{L^2}}\sum_{j\,
{\rm odd}}e^{ikj}f_j)$.

Linearizing around $k=\pi/2+k'$, we arrive at the continuum hamiltonian
\begin{equation}
H=-\int dk' \psi_{\alpha}^{\dagger}(k')\sigma_1 k'\psi_{\alpha}(k'),
\end{equation}
where $ \psi_{\alpha}=\left( \begin{array}{c} f_{\alpha e}\\
f_{\alpha o} \end{array} \right) $ and $\sigma_1=
\left( \begin{array}{cc} 0 & 1\\
1 & 0 \end{array}\right)$ is a Pauli matrix.  This is just
the hamiltonian of free Dirac fermions (we have set the velocity of
the fermions =1).
The corresponding (Euclidean space) lagrangian is
\begin{equation}
L= \bar{\psi}_{\alpha}\gamma_{\mu}\partial_{\mu}\psi_{\alpha},
\end{equation}
where $\bar{\psi}=\psi^{\dagger}\gamma_0$,
and $\mu=0,1$, and the $\gamma$ matrices
are
\begin{equation}
\gamma_0=\sigma_3,\,\,\, \gamma_1=-\sigma_2.
\end{equation}
In 1+1D, we define $\gamma_5$ matrix as $\gamma_5=
-i\gamma_0\gamma_1=\sigma_1$,
which has the property
\begin{equation}
\{\gamma_5,\gamma_{\mu}\}=0,\,\,\,\epsilon_{\mu\nu}\gamma_{\nu}=
i\gamma_{\mu}\gamma_5,
\end{equation}
($\epsilon_{\mu\nu}$ is the antisymmetric tensor with $\epsilon_{01}=1$).

Including the fluctuations around the mean field (the
fluctuations of $\lambda_i$  and the phase of $\chi_{ij}$) amounts to
coupling the fermions to
a U(1) gauge field by the minimal prescription.  Hence the continuum version
of \{the mean field + fluctuations\} is the two flavor Schwinger model
\begin{equation}
L= \bar{\psi}_{\alpha}\gamma_{\mu}(\partial_{\mu}-ia_{\mu})\psi_{\alpha}.
\end{equation}
The apparent gauge coupling (bare coupling)
 is infinitely strong, as there is no
kinetic term for the gauge field.

 Above lagrangian is obviously invariant
under the global SU(2) transform (spin rotation symmetry)
\begin{equation}
\psi_{\alpha}\rightarrow
(\exp(i\phi^l\tau^l))_{\alpha\beta}\psi_{\beta}
\end{equation}
where $\tau^l, l=1,2,3$ are Pauli matrices (belonging to a space different
from that of $\sigma_l$).
 The lagrangian is also invariant under the
``chiral transformation''
\begin{equation}
\psi_{\alpha}\rightarrow \exp(-i\theta\gamma_5)
\psi_{\alpha}.
\label{chiral}
\end{equation}
This ``chiral symmetry'' is {\it explicitly} broken by higher
derivative terms ignored in taking the continuum limit, like
\begin{equation}
L'= \bar{\psi}_{\alpha}\gamma_5\gamma_{\mu}(\partial_{\mu}-
ia_{\mu})^2\psi_{\alpha}.
\label{higher}
\end{equation}
If we label the states by the left and right movers,
$
f_R \approx f_k\,\,  , \,\, f_L \approx f_{k-\pi}
$,
where $0 < k < \pi$, the chiral transformation corresponds to
$
f_R \rightarrow f_R e^{-i\theta} \,\, , \,\, f_L \rightarrow f_Le^{i\theta}.
$

We now consider spin correlation functions at the mean field level ({\it i.e.}
ignoring gauge fields).
The spin operators  in the continuum  has two
contributions (uniform \& staggered):
\begin{eqnarray}
S^l(x_1)\approx [f_{\alpha e}^{\dagger}(x_1)\frac{\tau_{\alpha\beta}^l}{2}
f_{\beta e}(x_1)
+f_{\alpha o}^{\dagger}(x_1)\frac{\tau^l_{\alpha\beta}}{2}
f_{\beta o}(x_1)]\nonumber\\
+(-1)^{x_1}[f_{\alpha e}^{\dagger}(x_1)\frac{\tau_{\alpha\beta}^l}{2}
f_{\beta e}(x_1)
-f_{\alpha o}^{\dagger}(x_1)\frac{\tau^l_{\alpha\beta}}{2}
f_{\beta o}(x_1)]\nonumber\\
= \bar{\psi}_{\alpha}(x_1)\gamma_0\frac{\tau^l_{\alpha\beta}}{2}
\psi_{\beta}(x_1)
+(-1)^{x_1}\bar{\psi}_{\alpha}(x_1)\frac{\tau^l_{\alpha\beta}}{2}
\psi_{\beta}(x_1).
\end{eqnarray}
To evaluate the
spin correlation function
\begin{eqnarray}
\langle S^+(x)S^-(0) \rangle &=&
\langle\bar{\psi}\gamma_0\tau^{+}\psi(x)\bar{\psi}
\gamma_0\tau^{-}\psi(0)\rangle
+ (-1)^{x_1}\langle\bar{\psi}\tau^{+}\psi(x)\bar{\psi}\tau^{-}\psi(0)\rangle
\nonumber \\
&=&
 \langle\bar{\psi}_1\gamma_0\psi_2(x)\bar{\psi}_2\gamma_0\psi_1(0)\rangle
+ (-1)^{x_1}\langle\bar{\psi}_1\psi_2(x)\bar{\psi}_2\psi_1(0)\rangle
\label{spincor}
\end{eqnarray}
($\tau^{\pm}=(\tau^1\pm i\tau^2)/2$),
we need the fermion Green's function
\begin{equation}
{\bf G}_{\alpha\beta}(x)=
\langle\psi_{\alpha}(x)\bar{\psi}_{\beta}(0)\rangle\equiv
G(x)\delta_{\alpha\beta}
\end{equation}
which can be obtained from the momentum space Green function $G(k)
=-ik\gamma/k^2$ as
\begin{equation}
G(x)
=\gamma_{\mu}\frac{\partial}{\partial x_{\mu}}\int   \frac{d^2k}{(2\pi)^2}
\frac{e^{-ik\cdot x}}{k^2}=
-\frac{x\gamma}{2\pi x^2}.
\end{equation}
Here and from now on, unless otherwise specified, we use the usual
field theory notation: $k=(k_0,k_1), x=(x_0,x_1)$ [italics denote
space time vectors];
$x\gamma\equiv x_{\mu}\gamma_{\mu}$;
$x^2=x_0^2+x_1^2$, etc.
Using Wick's theorem, we have
\begin{eqnarray}
\langle S^+(x)S^-(0)\rangle
&=&-{\rm tr}_{\gamma,\tau}[{\bf G}(x)\gamma_0\tau^{+}
{\bf G}(-x)\gamma_0\tau^{-}] 
- (-1)^{x_1}{\rm tr}_{\gamma,\tau}[{\bf G}(x)\tau^{+}
{\bf G}(-x)\tau^{-}]\nonumber\\
&=&\frac{1}{2\pi^2}\left[\frac{x_0^2-x_1^2}{(x_0^2+x_1^2)^2}+
(-1)^{x_1}\frac{1}{x_0^2+x_1^2}\right]
\nonumber \\
&=& \frac{1}{4\pi^2}\left[\frac{1}{x_{-}^2}+\frac{1}{x_{+}^2}
+(-1)^{x_1}\frac{2}{x_{-}x_{+}}\right], \nonumber\\
\end{eqnarray}
($x_{\pm}\equiv x_0\pm i x_1$;
${\rm tr}_{\gamma,\tau}$ denotes trace
over both the $\gamma$ and $\tau$ spaces (spinor and spin spaces)),
which does (and should) equal
the $\langle S_z(x)S_z(0)\rangle$ correlation function in the XY
model\cite{tsvelik}.
The equal time correlation function
$\langle{\bf S}(x_1)\cdot {\bf S}(0)\rangle$
behaves as
\begin{equation}
\langle{\bf S}(x_1)\cdot {\bf S}(0)\rangle=
\frac{3}{4\pi^2(x_1)^2}((-1)^{x_1}-1),
\end{equation}
(the spins on the same sublattice are not correlated at all,
while the correlation among spins on different sublattices are decaying
algebraically as $1/x_1^2$).
This
peculiar behavior, which was derived by
 Arovas and Auerbach in the  lattice version of
the fermionic mean field theory\cite{arovas} and agrees with
Bulaevskii's  Hartree-Fock treatment of the Jordan-Wigner
fermionized  Heisenberg model\cite{bule}, is viewed as a pathology
of the mean field theory: {\it we have lost a substantial amount of
antiferromagnetic correlation}.

\subsection{Schwinger model}
We now consider the effect of gauge fluctuations.  It is natural to expect
that the inclusion of gauge fluctuations will improve the mean field
picture.  The time component of the gauge field can be regarded to originate
from the Lagrange multiplier field (for no-double-occupancy); this corresponds
to  Gutzwiller-projected (half-filled tight binding)
 Fermi surface, which is known to be a
 pretty good description of 1d antiferromagnet\cite{kaplan,exactc}.

As mentioned earlier, our theory with fluctuations  is a Schwinger
model.
For reasons that will become clear shortly, we consider a slightly more general
case of $N$-flavors:
\begin{eqnarray}
Z&=&\int D\bar{\psi}D\psi Da_{\mu} \exp(-S),\nonumber \\
 S&=&\int d^2x \sum_{\alpha=1}^{N} \bar{\psi}_{\alpha}\gamma_{\mu}
(\partial_{\mu}
-ia_{\mu})\psi_{\alpha} + \frac{1}{4e^2}F_{\mu\nu}^2,\,\,\,\,\,\,
 (e^2=\infty).
\end{eqnarray}
The physical case is $N=2$; general (even) $N$  corresponds to an SU($N$)
antiferromagnet.

Integrating out the fermions gives
\begin{equation}
Z=\int  Da_{\mu} \exp(N\,{\rm Tr}\ln(1-i{\cal G}\gamma_{\mu}a_{\mu}))
\end{equation}
where ${\cal G}(x,x')=(\gamma_{\mu}\partial_{\mu})^{-1}\delta(x-x')$,
 and Tr denotes traces over the spinor space
and the position space (Tr=tr$\int d^2xd^2x'...$).
 The logarithm can be expanded, giving
\begin{equation}
Z =\int  Da_{\mu} \exp\left(-\frac{1}{2}\int d^2xd^2x'
a_{\mu}(x)\Pi_{\mu\nu}(x-x') a_{\nu}(x')\right),
\label{exact}
\end{equation}
where $\Pi_{\mu\nu}(x)= -N\,{\rm tr}[G(x)\gamma_{\mu}G(-x)\gamma_{\nu}]$.
Note that the beyond-Gaussian terms
(like $\Gamma_{\mu\nu\rho\delta}a_{\mu}a_{\nu}a_{\rho}a_{\delta}$)
are all  zero; the proof can
be found, for example, in Ref.\onlinecite{frishman}.

The polarization function $\Pi_{\mu\nu}(q)$ (in the momentum
space) contains a divergence that has to be regulated
using gauge invariant  schemes, like the dimensional regularization or
the Pauli-Villars regularization.  Relegating the details to Appendix A,
we have
\begin{equation}
\Pi_{\mu\nu}(k)=\frac{N}{\pi}\left(\delta_{\mu\nu}-
\frac{k_{\mu}k_{\nu}}{k^2}\right),
\label{polfermi}
\end{equation}
which means that the
 gauge boson acquires an infinite mass ($= e\sqrt{N/\pi}$).
The transversality of
Eq.\ref{polfermi} guarantees the conservation of the
current $j_{\mu}=\bar{\psi}_{\alpha}\gamma_\mu\psi_{\alpha}$:
\begin{equation}
q_{\mu}j_{\mu}(q)= q_{\mu}(i\Pi_{\mu\nu}(q)a_{\nu}(q))=0\,\,\, \rightarrow
\partial_{\mu}
j_{\mu}=0.
\end{equation}
On the other hand the current $j_{5\mu}
(=i\bar{\psi}_{\alpha}\gamma_5\gamma_{\mu}
\psi_{\alpha}=-\epsilon_{\mu\nu}j_{\nu})$ associated with the chiral symmetry
(Eq.\ref{chiral}) is
not conserved:
\begin{equation}
q_{\mu}j_{5\mu}=-q_{\mu}\epsilon_{\mu\nu}i\Pi_{\nu\rho}a_{\rho}
=-\frac{i N}{\pi}\epsilon_{\mu\nu}q_{\mu}a_{\nu}
  \rightarrow \,
\partial_{\mu}j_{5\mu}= -\frac{iN}{2\pi}\epsilon_{\mu\nu}F_{\mu\nu}.
\label{axial}
\end{equation}
This result, the so-called axial anomaly, can be regarded as either
a consequence
of or a condition for gauge invariance.  Equation (25) is in fact familiar
to the solid state
physicists\cite{Peskin}.  In an electric field, the equation of motion
for the crystal momentum
state is
\begin{equation}
\frac{d{\bf k}}{dt} = {\bf E}.
\end{equation}
In one dimension an electric field causes a shift of the occupation
between left moving and
right moving states.  Thus the left and right movers are in fact connected
and their densities are
not separately conserved.  It is easy to see that Eq.(26) is consistent
with Eq.(25).

The exact spin correlation functions can be evaluated with the
use of ``chiral rotation''\cite{lowenstein,fujikawa,stone,zinnjus}.
In this approach,
the gauge field is written as the sum of a div-free part
and a curl-free part:
\begin{equation}
a_{\mu}=\epsilon_{\mu\nu}\partial_{\nu}\theta_a + \partial_{\mu}\theta_b.
\label{chiralgauge}
\end{equation}
The transform
\begin{mathletters}
\begin{eqnarray}
\psi_{\alpha} &\rightarrow & \psi_{\alpha}'=\exp(-\gamma_5 \theta_a -
i\theta_b)\psi_{\alpha} \label{Eq.(28a)} \\
\bar{\psi}_{\alpha} &\rightarrow & \bar{\psi}_{\alpha}'=
\bar{\psi}_{\alpha}\exp(-\gamma_5 \theta_a + i\theta_b) \label{Eq.(28b)}
\end{eqnarray}
\end{mathletters}
decouples the gauge field from the $\psi'$  fermions.
We note that corresponding to Eq.(28b), we have
\begin{equation}
\psi^\dagger_\alpha \rightarrow \psi^{\prime\dagger}_\alpha \exp
(\gamma_5\theta_a + i\theta_b)
\end{equation}
because $\gamma_0$ and $\gamma_5$ anticommute.  Thus
$\psi^\dagger_\alpha\psi_\alpha =\psi'^\dagger_\alpha\psi_\alpha' $ and the
transformation is unitary.  Here $\psi_\alpha$ and $\psi^\dagger_\alpha$
are treated as
independent Grassmannian variables and $\exp(\gamma_5\theta_a)$ should not
be thought of as an
amplitude transformation.  Alternatively, we point out that in real time
(as opposed to imaginary
time used here), the transformation to cancel the $a_\mu$ field will indeed
be a phase rotation and
takes the form
$
\psi_\alpha \rightarrow \psi^\prime_\alpha \exp
(-i\gamma_5\theta_a + i\theta_b)
$.
That this must be the case is evident from Eq.(27), because $a_0 =
\partial_1\theta_a$ changes
from real to imaginary upon going from Minkowski to Euclidean space.
Unlike the usual gauge
transformation, $ia_0 f_L$ is cancelled by $\partial_1 f_L$ and $ia_1f_L$
is cancelled by
$\partial_0 f_L$ under the chiral rotation, so that $\theta_a$ must also
change from real to imaginary.

Even though the chiral rotation is unitary, it does not leave the Grassmann
measure invariant, due
to the fact that $f_R$ and $f_L$ are not truly independent, as noted earlier.
The jacobian $\exp(J)$
for the change of measure
\begin{equation}
D\psi D\bar{\psi}=  D\psi'D\bar{\psi}' \exp(J)  \nonumber
\end{equation}
can be found easily using the axial anomaly condition Eq.\ref{axial}(see
Appendix B for details):
\begin{equation}
J=-\frac{N}{2\pi}\int d^2x (\partial_{\mu}\theta_a)^2.
\end{equation}
Alternative derivation can be found in Ref.\onlinecite{zinnjus}.
In terms of the new fields, the functional
integral is
\begin{equation}
Z\!=\!\int\! D\bar{\psi}'D\psi'D\theta_a
\exp -\!\int\!\! d^2x\!
\left(\bar{\psi}_{\alpha}'\gamma_{\mu}\partial_{\mu}\psi_{\alpha}'+
\frac{N}{2\pi}
(\partial_{\mu}\theta_a)^2\right).
\end{equation}
These are ``free'' fields,
and now the spin correlation function (Eq.\ref{spincor})
can be evaluated easily:
\begin{eqnarray}
\langle\bar{\psi}_1\gamma_0\psi_2 (x) \bar{\psi}_2\gamma_0\psi_1(0)\rangle&=&
\langle\bar{\psi}_1'\gamma_0\psi_2' (x)
\bar{\psi}_2'\gamma_0\psi_1'(0)\rangle \nonumber\\
&=& \frac{1}{2\pi^2}\frac{x_0^2-x_1^2}{(x_0^2+x_1^2)^2},\nonumber\\
\langle\bar{\psi}_1\psi_2 (x) \bar{\psi}_2\psi_1(0)\rangle
&=&\langle\bar{\psi}_1'e^{2\gamma_5\theta_a}\psi_2' (x) \bar{\psi}_2'
e^{2\gamma_5\theta_a}\psi_1'(0)\rangle\nonumber\\
&=&
\frac{1}{2\pi^2x^2}e^{2\langle(\theta_a(x)-\theta_a(0))^2\rangle}\nonumber\\
&=&\frac{{\cal C}}{x^2}e^{\frac{1}{N}\ln(x^2)}=
\frac{{\cal C}}{(x^2)^{1-1/N}}, \label{thirtythree}
\end{eqnarray}
where we have used
\begin{eqnarray}
\langle(\theta_a(x)-\theta_a(0))^2\rangle &=& -\frac{2\pi}{N}
\int^{\Lambda} \frac{d^2k}{(2\pi)^2} \frac{1}{k^2}
(e^{ik\cdot x}-1) \nonumber \\
&=&\frac{1}{2N}\ln(x^2\Lambda^2),
\end{eqnarray}
($\Lambda$ is a UV cutoff originating from the lattice theory, and
${\cal C}$
is a nonuniversal constant that depends on high energy details $\Lambda$).
In the physical case ($N=2$),  we then have
\begin{equation}
\langle{\bf S}(x)\cdot{\bf S}(0)\rangle\sim (-1)^{x_1}
\frac{1}{\sqrt{x_1^2+x_0^2}}
\label{simple}
\end{equation}
which agrees with the more accurate result\cite{boso}
up  to a  $\ln^{1/2}(x^2)$ factor.
The log factor, not captured by the Schwinger model,
must be due to terms ignored in our derivation from the lattice theory,
{\it e.g.}, the amplitude fluctuation of the RVB field;
this is  analogous to
the bosonization theory of Heisenberg model, in which the Umklapp processes
give rise to logarithmic correction (prefactor) to the power
law\cite{lutherpeschel}.
The simple correlation function of Eq.\ref{simple} is
 known to  capture the low energy (temperature) properties of
the Heisenberg spin chains quite well\cite{sachdev}.

\subsection{additional remarks}
Before moving on to
the perturbative treatment of the same theory, we briefly compare the gauge
theory
approach with other approaches.  Perhaps the best known treatment of the
 Heisenberg spin chain relies on the Jordan-Wigner transformation
($f_i= e^{i\phi_i}S_i^{-}$, \,$\phi_i=\pi\sum_{j=1}^{i-1} S_i^{+}S_i^{-}$).
However, unlike the X-Y model, this does not lead to a theory
of free fermions, but to a theory with a 4-fermion interaction with coupling
constant of order unity, which is then treated by bosonization
methods\cite{lutherpeschel}. A drawback of the Jordan-Wigner approach is that
the SU(2) spin symmetry is easily lost, and the correct exponent ($=-1$) of
the staggered spin correlation function has to be determined rather
indirectly.

Haldane\cite{hald1d}, on the other hand, proposed to analyze the Hubbard model,
 exploiting the ``equivalence''
 between the  spin sector of the 1d Hubbard model
\begin{equation}
H=-\frac{1}{2}(JU)^{1/2}\sum_{i\sigma}(c_{i\sigma}^{\dagger}c_{i+1,\sigma}+
{\rm h.c.})
+U\sum_i (n_{i\uparrow}-\frac{1}{2})(n_{i\downarrow}-\frac{1}{2})
\end{equation}
and the 1d Heisenberg model, that holds even in the weak coupling limit.
The Hubbard model can be treated by a bosonization method\cite{hald1d,tsvelik}
 that respects the SU(2)
symmetry at all stages, and the correct exponent of the spin correlation
function can be obtained directly.  We can easily adapt Haldane's approach
to the Schwinger model,
which as we point out, is equivalent to the RVB mean field theory plus
gauge fluctuations.  This
line of approach was recently given in ref.[24].  Instead of treating the
gauge fluctuation by
chiral rotation, we first integrate out the gauge field in Eq.(20), which
simply enforces the
constraint of no double occupation, i.e., the charge fluctuation is
completely suppressed.  Now we
can treat the fermion problem by bosonization in the standard way.  We write
\begin{equation}
(f_{sR}(x_1),f_{sL}(x_1))=
e^{i\sqrt{\pi} \phi_s(x_1)}(e^{i\sqrt{\pi} \theta_s(x_1)
  + ik_Fx_1},\,\, e^{-i \sqrt{\pi} \theta_s(x_1) - ik_Fx_1})
\end{equation}
for $s$ = spin up or down, and
$
\theta_\uparrow = \frac{1}{\sqrt{2}} \left( \theta_\rho + \theta_\sigma \right)
$,
$
\theta_\downarrow = \frac{1}{\sqrt{2}}
\left( \theta_\rho - \theta_\sigma  \right)
$,
$
\phi_\uparrow =  \frac{1}{\sqrt{2}}\left( \phi_\rho + \phi_\sigma \right)
$,
$
\phi_\downarrow =\frac{1}{\sqrt{2}}\left( \phi_\rho - \phi_\sigma \right)
$,
where $\partial_{x_1}\theta_{\rho (\sigma)}$ is proportional to the charge
(spin) density and is
the conjugate variable to $\phi_{\rho(\sigma)}$.  In the effective
Lagrangian, spin and charge
fluctuations separate.  For free fermions $\theta_\sigma$ and $\theta_\rho$
contribute equally to
the $2k_F$ spin-spin correlation functions, yielding Eq.(19).  In our case
the charge fluctuation
is suppressed and we may ignore the $\theta_\rho$ degrees of freedom.  The
$2k_F$ spin correlation function decays with an exponent exactly one half 
than that of the free fermion, and we recover Eq.(\ref{thirtythree}).
This line of reasoning has a slight advantage over Haldane's original
treatment in that the energy scale $J$ for the fermion bandwidth is
correctly produced.


\subsection{perturbation theory}
We now examine the gauge theory in terms of
the perturbation theory in $1/N$.
  The key point is that the nature
of the
perturbative correction to the mean field results is very different for
the uniform part and the antiferromagnetic (staggered) part of the
spin correlation: While ${\bf Q}=\pi$ response is strongly affected by
perturbative correction,
the ${\bf Q}=0$ response receives no correction at
all (the absence of any correction is a special feature of the 1d).

\begin{figure}
\epsfxsize=\columnwidth
\epsfbox{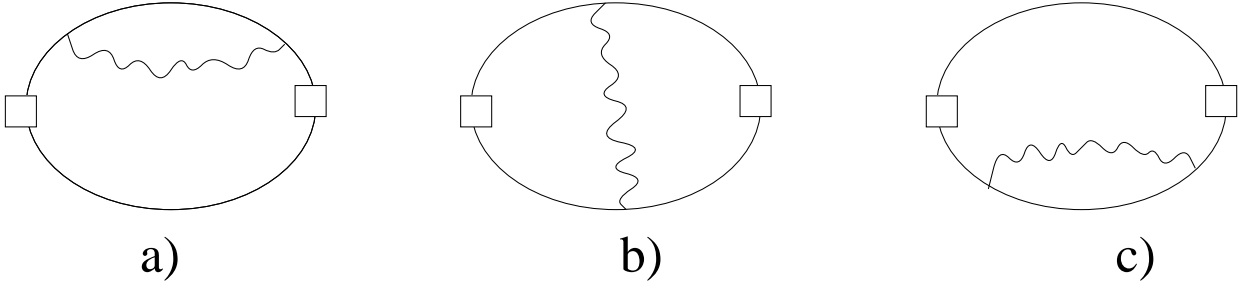}
\caption{Leading $1/N$ correction to uniform spin correlation.}
\label{1dunif}
\end{figure}

\begin{figure}
\epsfxsize=\columnwidth
\epsfbox{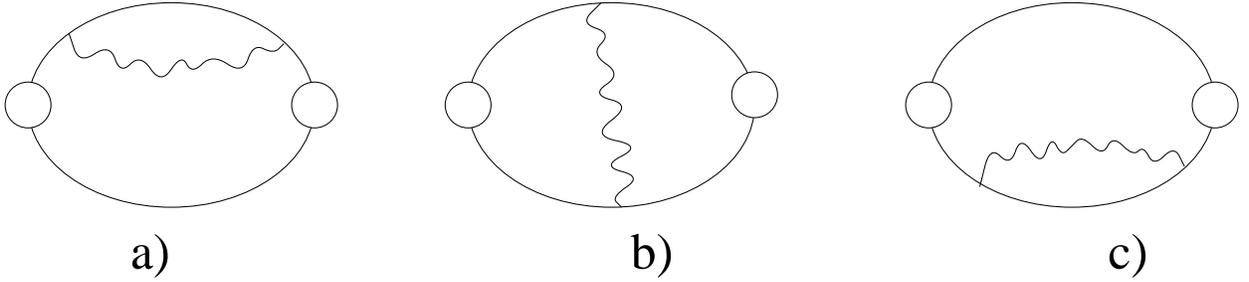}
\caption{Leading $1/N$ correction to staggered spin correlation.}
\label{1dstag}
\end{figure}

The leading $1/N$ correction to the spin correlation functions
can be straightforwardly evaluated.
 They are represented by the Feynman diagrams in Fig.\ref{1dunif}
 and Fig.\ref{1dstag}.
Within the usual Faddeev-Popov scheme of gauge fixing (the introduction
of  the $\frac{1}{2\lambda} (\partial\cdot a)^2$ term to the lagrangian),
the gauge propagator (represented by the wiggly line) is given by
\begin{equation}
D_{\mu\nu}=\langle a_{\mu}a_{\nu}\rangle=\frac{\pi}{N}
\left(\delta_{\mu\nu}-\frac{q_{\mu}q_{\nu}}{q^2}\right)+\lambda
\frac{q_{\mu}q_{\nu}}{(q^2)^2}.
\end{equation}
We choose the Landau gauge ($\lambda=0$) which is a natural
choice, since in this gauge no infrared divergence occurs in
the perturbation theory.
The fermion propagator $G(p)=(ip\gamma)^{-1}$
is represented by a solid line.  The fermion-gauge
vertex is simply $i\gamma_{\mu}$; the external current vertex is
$\gamma_0$ for the uniform part (represented by a square) and
1 for the staggered part (represented by a circle).  Of course
a trace is taken over the fermion loops.

\narrowtext

\narrowtext

In 1+1D, the transverse projector has a special property
\begin{equation}
\delta_{\mu\nu}-\frac{q_{\mu}q_{\nu}}{q^2}
=\epsilon_{\mu\rho}\epsilon_{\nu\delta}
\frac{q_{\rho}q_{\delta}}{q^2}
\end{equation}
which, together
with the ``Ward identity''
\begin{equation}
G(p+q)q\gamma G(p)=i(G(p) - G(p+q)),
\end{equation}
simplifies the algebra substantially (Note $\gamma_{\mu}\epsilon_{\mu\rho}
p_{\rho}=-ip\gamma\gamma_5$).  For example, the diagram \ref{1dunif}b is
\begin{eqnarray}
\frac{\pi}{N}\int \frac{d^2p}{(2\pi)^2}\frac{d^2q'}{(2\pi)^2}
 \,{\rm tr} [G(p+q)q'\gamma\gamma_5
G(p+q+q')\gamma_0
G(p+q')q'\gamma\gamma_5G(p)\gamma_0]/q'^2 \nonumber\\
=- \frac{\pi}{N}\int \frac{d^2p}{(2\pi)^2}\frac{d^2q'}{(2\pi)^2}
{\rm tr} [(G(p+q)-
G(p+q+q'))\gamma_0
(G(p+q')-G(p))\gamma_0]/q'^2.\nonumber \\
\label{vertexc}
\end{eqnarray}
It's straightforward to show that sum of the diagrams \ref{1dunif}a+c is the
same as Eq.\ref{vertexc}, except for a minus sign.  Therefore in the uniform
channel, the vertex correction and the self energy correction cancel.
Similar cancellation is expected at all orders of  perturbation theory; the
nonrenormalization of uniform part of the spin correlation function
is quite natural,
since in our theory
\begin{equation}
\langle\bar{\psi}_1\gamma_0\psi_2\bar{\psi}_2\gamma_0\psi_1\rangle
\propto\langle j_0j_0\rangle=\Pi_{00},
\end{equation}
and Eq.\ref{exact} is an exact result.

On the other hand, the diagrams in  the staggered channel do not cancel.
The sum of the diagrams \ref{1dstag}a and \ref{1dstag}c are equal to
\ref{1dstag}b, which is given by
\begin{eqnarray}
\frac{\pi}{N}\int \frac{d^2p}{(2\pi)^2}\frac{d^2q'}{(2\pi)^2}
\, {\rm tr} [(G(p+q)-
G(p+q+q'))1
(G(p+q')-G(p))1]/q'^2\nonumber \\
=\frac{2\pi}{N}\int \frac{d^2p}{(2\pi)^2}\frac{d^2q'}{(2\pi)^2}\,
 {\rm tr} [G(p+q')G(p+q)
-G(p+q)G(p))]/q'^2.
\end{eqnarray}
Therefore, in the coordinate space, the $1/N$ correction is
\begin{equation}
{\rm tr}[G(x)G(-x)]\frac{4\pi}{N}
\int \frac{d^2q}{(2\pi)^2} \frac{1}{q^2}(e^{iq\cdot x}-1)
=\frac{1}{2\pi^2 N x^2}\ln(x^2).
\end{equation}
Similar (but a lot more tedious) calculation
 would  show  that $1/N^2$ correction is
given by $\frac{1}{4\pi^2N^2x^2}\ln(x^2)$.  In other words, the
perturbation series exponentiates:
\begin{equation}
\langle\bar{\psi}_1\psi_2(x)\bar{\psi}_2\psi_1(0)\rangle=
\frac{1}{2\pi^2x^2}(1+
(1/N)\ln(x^2)
+\frac{1}{2}(1/N)^2\ln^2(x^2)+...)\propto \frac{1}{(x^2)^{1-1/N}},
\end{equation}
giving the same result obtained from the chiral rotation approach.

\section{Two-dimensional Undoped Cuprates}

The success of 1+1D
 gauge theory with Dirac fermions in describing the Heisenberg
spin chain
tempts us that a similar theory of massless Dirac fermions strongly coupled to
a U(1) gauge field might describe a 2d quantum antiferromagnet.  In fact, it is
known from lattice gauge theories\cite{diamantini} that this is indeed so.
In connection with  high \Tc cuprates,
Marston\cite{marston}, Laughlin\cite{laughzou},
 and others have noted that
the ``chiral symmetry breaking'' in 2+1D U(1) gauge theory, discussed by
Pisarski\cite{pisarski} and Appelquist {\it et al.}\cite{appelquist}
in particle physics context, corresponds to the N\'{e}el ordering.
In this section, we shall discuss this matter carefully, and
clarify some issues related to the pattern of symmetry breaking and
Goldstone bosons (the particle physics literature\cite{pisarski,appelquist}
 envisions symmetry breaking pattern U(4)$\rightarrow {\rm U(2)\bigotimes
 U(2)}$, hence
$16-2\cdot 4$= 8 Goldstone bosons, while the symmetry breaking pattern for
N\'{e}el ordering is SU(2)$\rightarrow$
U(1) which gives $3-1=2$ Goldstone bosons).
SU(2)  gauge theories with massless Dirac fermions may also describe
the quantum antiferromagnet\cite{affzou,mudry2}, but we
shall not consider this possibility because of the greater complexity of the
nonabelian gauge theories.

\subsection{Dirac fermions and the 2d Heisenberg antiferromagnet}
A 2+1D theory of Dirac fermions
\begin{equation}
L=\bar{\psi}_{\alpha}\partial_{\mu}\gamma_{\mu}\psi_{\alpha},
\label{2dmf}
\end{equation}
($\mu$=0,1,2) contains fermions whose density
of states  behaves as $\sim |\epsilon|$
(in a general $D=d+1$ dimensions, the density of states will be
$\sim |\epsilon|^{d-1}$).
 In condensed matter context, the paramagnetism of
such fermions will result in the uniform susceptibility behaving as
$T^{d-1}$.   This seems to have
 little in common with the 2d antiferromagnet, but
we shall see that a  gauge field coupled to fermions can produce the
correspondence with the physics of 2d antiferromagnet.
Such a picture  can be motivated from the ``$\pi$-flux'' phase mean field
ansatz of the Heisenberg hamiltonian. The ansatz is so named since the
 phase of the product of the mean field parameter $\chi_{ij}$
around each plaquette
(Im$\ln(\chi_{12}\chi_{23}\chi_{34}\chi_{41})$) is $\pi$.
Despite the ``flux,''
the mean field does not break the parity and time reversal symmetry
since the flux of $\pi$ is equal to $-\pi$.  This phase has  a lower
energy than the BZA phase (with a large Fermi surface)\cite{bza}; it also has
the fermion spectrum
\begin{equation}
\epsilon({\bf k})=v\sqrt{\cos^2(k_x)+\cos^2(k_y)}
\end{equation}
that roughly captures the high energy features of the undoped
cuprates and the dispersion of a single hole in the
antiferromagnet\cite{mitstanford,laughlin}.
The low energy fermionic excitations of the flux phase
 reside near two ``Fermi points'',
${\bf k}_{1,2}=(\pi/2,\pm \pi/2)$.  Linearizing around these points
gives the continuum theory\cite{flux}
\begin{equation}
L=\bar{\psi}_{\alpha a}\partial_{\mu}\gamma_{\mu}\psi_{\alpha a},
\end{equation}
where
$\psi_{\alpha 1}=\left( \begin{array}{c} f_{\alpha 1e} \\
f_{\alpha 1o} \end{array} \right)$,
$\psi_{\alpha 2}=\left( \begin{array}{c} f_{\alpha 2o} \\
f_{\alpha 2e} \end{array}\right)$,
($a=1,2$ labels the two Fermi points; $e$,$o$ denote even and odd sites.)
Organizing the  $\psi_{\alpha 1},\psi_{\alpha 2}$ fields into a single spinor
$\psi_{\alpha}\equiv \left( \begin{array}{c} \psi_{\alpha 1}\\ \psi_{\alpha 2}
\end{array}\right)$,
we have a theory described by the lagrangian of Eq.\ref{2dmf}, with
 $4\times 4$ $\gamma$-matrices:
\begin{equation}
\gamma_0=\left( \begin{array}{cc} \sigma_3 & 0\\0 &-\sigma_3 \end{array}\right),
\,\,\,
\gamma_1=\left( \begin{array}{cc} \sigma_2 & 0\\0 & -\sigma_2
\end{array}\right),
\,\,\,
\gamma_2=\left( \begin{array}{cc} \sigma_1 & 0\\0 & -\sigma_1
\end{array}\right).
\end{equation}
Similarly to the 1d case, the uniform spin $S_{{\bf Q}=0}^l$ is given by
$\bar{\psi}_{\alpha}\gamma_0\tau_{\alpha\beta}^l\psi_{\beta}$, while
the staggered spin $S_{{\bf Q}=(\pi,\pi)}^l$ is given by
$\bar{\psi}_{\alpha}\tau_{\alpha\beta}^l\psi_{\beta}$.  The spin correlation
function
$\langle S^+(x)\cdot S^-(0)\rangle $
can be evaluated easily at the mean field level using the Green's function
\begin{equation}
{\bf G}_{\alpha\beta}(x)=\delta_{\alpha\beta}G(x)
=
\delta_{\alpha\beta}\gamma_{\mu}\frac{\partial_{\mu}}
{\partial x_{\mu}}\int \frac{d^3q}{(2\pi)^3} \frac{e^{-iq\cdot x}}{q^2}=
\delta_{\alpha\beta}\frac{x\gamma}{4\pi(x^2)^{3/2}}
\end{equation}
(as in 1+1D, the italics like $x$, $q$ denote space-time vectors, and
$q^2=q_0^2+{\bf q}^2, x^2=x_0^2+{\bf x}^2$.)
We have,
\begin{eqnarray}
\langle S^{+}(x)S^{-}(0)\rangle&=& -{\rm tr}[G(x)\gamma_0 G(-x)\gamma_0]
- (-1)^{x_1+x_2}
{\rm tr}[G(x)G(-x)] \nonumber \\
&=& \frac{1}{4\pi^2}\left[\frac{x_0^2-{\bf x}^2}{x^6} +
(-1)^{x_1+x_2}\frac{1}{x^4}\right].
\end{eqnarray}
The equal time correlation function is then
\begin{equation}
\langle S^{+}(0,{\bf x})S^{-}(0,0)\rangle=
\frac{(-1)^{(x_1+x_2)}-1}{4\pi^2{\bf x}^4}.
\label{twodcorr}
\end{equation}
The mean field spin correlation function falls off as $1/{\bf x}^4$, and
again as in 1d, the spins on the same sublattice are not
correlated\cite{laughzou,arovas}.  Note, however, that due to our neglect
of $(\pi,0)$ contributions, Eq.\ref{twodcorr} differs somewhat from the
more accurate expressions of Refs.\cite{laughzou,arovas}.

  To go beyond the mean field level,
gauge fluctuation is included by the minimal
coupling scheme,  leading to the following theory:
\begin{equation}
Z=\int\!D\bar{\psi}D\psi D a_{\mu}\, \exp\!\left(\!-\int\! d^3x \,
\!\sum_{\alpha=1}^N\bar{\psi}_{\alpha}(\partial_{\mu}-
ia_{\mu})\gamma_{\mu}\psi_{\alpha}\right)
\label{qed3}
\end{equation}
where $N$ is a general number (in the physical case, $N=2$).
  The absence of the
$\frac{1}{4e^2}F_{\mu\nu}^2$ term
means the bare coupling is infinite, but the theory is still  sensible:
 the infrared behavior of the theory is
well-behaved (within $1/N$ expansion, as we  shall see),
 and the original lattice theory sets a ultraviolet cutoff.
(Ultraviolet divergence can be also regulated by the the kinetic term
$\frac{1}{4\tilde{e}^2}F_{\mu\nu}^2$ with $\tilde{e}^2 < \infty$.)
As in 1+1D, the lagrangian contains more
symmetries than the Heisenberg model. For example the theory is invariant
under the transform $\psi_{\alpha}\rightarrow \exp(i\gamma_{4,5}\theta)
\psi_{\alpha}$ (where $\gamma_4=
\left( \begin{array}{cc}0 & I\\I &0 \end{array}\right),
\gamma_5=
\left( \begin{array}{cc}0 & I\\-I &0 \end{array}\right)$; $I$ is a $2\times 2$
unit matrix).
 Again, these symmetries are broken by  higher order terms (lattice
effects)  which have
been ignored.

\subsection{content of the gauge theory}
We now explore the physical content of the theory with massless Dirac
fermions.  Integrating out the fermions generates the dynamics for the
gauge field\cite{IL} (see Appendix A for details)
\begin{eqnarray}
Z&=&\int Da_{\mu} \exp\left(-\frac{1}{2}\int \frac{d^3q}{(2\pi)^3}\,
a_{\mu}(q)\Pi_{\mu\nu}(q) a_{\nu}(q)\right),\nonumber\\
\Pi_{\mu\nu}&=&\frac{N}{8}\sqrt{q^2}\left(\delta_{\mu\nu}-
\frac{q_{\mu}q_{\nu}}{q^2}\right).
\label{zeroth}
\end{eqnarray}
The form of $\Pi_{\mu\nu}$ indicates that, unlike the 1+1D case, the gauge
field is massless, though the infrared behavior is not as singular as
the free gauge field ({\it i.e.}, $L=\frac{1}{4}F_{\mu\nu}^2$).
The effect of the fermion-gauge field interaction can be analyzed
perturbatively in $1/N$.  The gauge propagator in the Faddeev-Popov
prescription is
\begin{equation}
D_{\mu\nu}(q)= \frac{8}{N\sqrt{q^2}}\left(\delta_{\mu\nu}-
\frac{q_{\mu}q_{\nu}}{q^2}\right)+\lambda \frac{q_{\mu}q_{\nu}}{q^4}.
\end{equation}
As in 1+1D (Sec.II), we choose the Landau gauge ($\lambda=0$)
 to avoid spurious
infrared divergences.

\begin{figure}
\epsfxsize=\columnwidth
\epsfbox{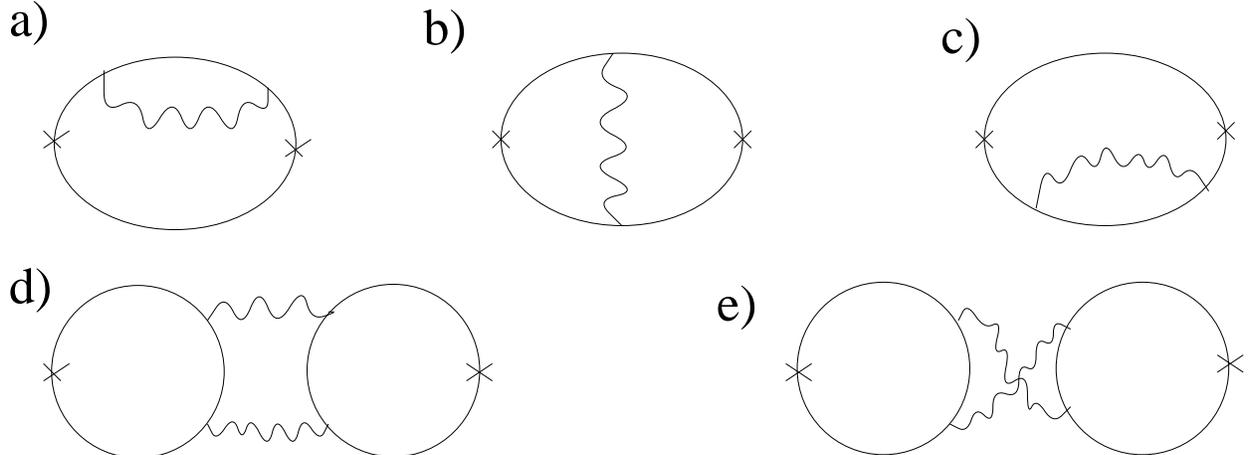}
\caption{Leading $1/N$ correction to vacuum poloarization.  The diagrams
d) and e) are zero due to Furry's theorem.}
\label{polar}
\end{figure}

  The fermion self-energy at the leading order
in $1/N$ is given by
\begin{equation}
\Sigma(k)=i\int \frac{d^3q}{(2\pi)^3}
 \gamma_{\mu}\frac{(k+q)\gamma}{(k+q)^2}\gamma_{\nu}D_{\mu
\nu}(q)  \sim i k\gamma\ln\left(\frac{\Lambda^2}{k^2}\right).
\end{equation}
This log divergence does {\it not} occur in the $1/N$ correction to
the polarization function represented by the diagrams in Fig.\ref{polar}.
These diagrams
have been calculated  explicitly in a different context by Chen
{\it et al.}\cite{chen}. They found that the logarithmic divergences in the
self energy correction (Figs.\ref{polar} a+c) are cancelled by the
vertex correction (Fig.\ref{polar} b); the diagrams sum  to
\begin{equation}
\Pi_{\mu\nu}^{1/N}(q) = \frac{3}{4\pi^2}\sqrt{q^2}\left(\delta_{\mu\nu}
-\frac{q_{\mu}q_{\nu}}{q^2}\right),
\end{equation}
which is of the same form as the zeroth order result (Eq.\ref{zeroth}) but
down by some factor involving $1/N$.
Thus, although we do not have a complete cancellation (like 1d),
the gauge
field is essentially unrenormalized, except for
some modification of the effective coupling constant ($1/N\rightarrow
1/N'\approx 1/N$). In other words, $Z_3$ (charge renormalization) $\approx
1$.  Weak
correction to the gauge propagator (despite strong self energy correction)
was seen in several other contexts, including the
half-filled Landau level (fermion
Chern-Simons theory)\cite{ybkim},
 bosonic Chern Simons theories\cite{semenoff}, and the uRVB gauge
theory\cite{PN}.    This robustness
is due to the fact that $\Pi_{\mu\nu}$ is a correlation function of
conserved currents,
and conserved currents have no anomalous dimensions
(as a consequence of the Ward identiy)\cite{wenscale}.
 The foregoing argument
provides some optimism for perturbation theory in $1/N$.

\narrowtext

Because the uniform spin correlation $\Pi_u(x)=
\langle\bar{\psi}_1\gamma_0\psi_2(x)\bar{\psi}_2\gamma_0\psi_1(0)\rangle
\sim \Pi_{00}(x)$,
it wouldn't be renormalized significantly.
On the other hand, the staggered part  $\Pi_s(x)$ of the spin
correlation function does not involve conserved currents, and
is  expected to be strongly affected by gauge fluctuations.  This
difference can be more or less seen in perturbation theory.
The diagramatic representation of $\Pi_u$ is
the same as in the 1+1D case (Fig.\ref{1dunif}).  The external vertices
in this case are a fermion-gauge field vertex ($\gamma_0$).
Using  the Ward identity
\begin{equation}
\frac{\partial}{\partial p_{\mu}}\Sigma(p)= i\Gamma_{\mu}(p,p),
\end{equation}
we can see that the 2 overlapping divergences of Fig.\ref{1dunif}b are
cancelled by  self-energy bubbles of
Fig.\ref{1dunif}a+c.
For $\Pi_s$, the external vertices are  1 (unit matrix in the spinor
space). In that case, the cancellation does not
occur, and divergences develop in $\Pi_s$.

The question is, what would be ultimately the behavior of $\Pi_s$?
Would $\Pi_s(x)$ be characterized by simple power law correlations
like $\Pi_s(x)\sim 1/(x^2)^{\alpha}, (\alpha < 2)$ like the 1d case?
This might be a realization of Anderson's 2d Luttinger liquid
scenario\cite{anderson2},
but we feel that {\it a priori} there is no reason to expect so;
after all, 1+1D is rather special, with all sort of fascinating
features
like the infinite dimension of the
conformal group\cite{tsvelik}
and  Coleman's theorem\cite{col_thrm} which prohibits
spontaneous symmetry breaking.  More plausibly, we would expect a
symmetry-broken phase
(N\'{e}el order) or a symmetric phase with a more complicated
magnetic correlations.

\subsection{spontaneous symmetry breaking}
The previous section has identified
 a possible antiferromagnetic instability, which corresponds to  the
 staggered magnetization $\langle\bar{\psi}\tau^l
\psi\rangle$ acquiring a definite orientation --- an SU(2) symmetry breaking.
We now examine this possibility more closely.
In the symmetry-broken case, the fermions
acquire a mass (dynamic mass generation).

\begin{figure}
\epsfxsize=\columnwidth
\epsfbox{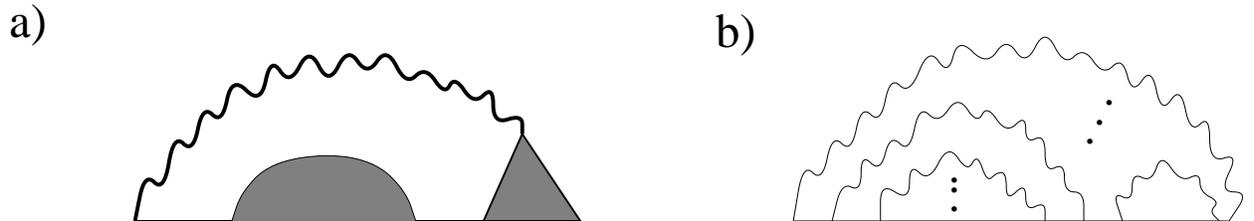}
\caption{Schematic representation of the Schwinger-Dyson equation. In part
a) the solid line with shaded blob 
 is the self-consistent Green's function of the fermions
 ${\bf G}$,
the thick wiggly line is the dressed gauge field (which incorporates the
changes in the vacuum polarization due to changes in fermion Green's 
function), and the shaded triangle is
the dressed vertex.  Part b) is a representation of the contribution
to the self energy.}
\label{sd}
\end{figure}

  Without loss of generality, we assume
that the rotation symmetry is broken in the z-direction.  Then the fermion
Green's function ${\bf G}(k)$ becomes
$1/(ik\gamma + m(k)\tau^3)$. Note that ${\bf G}$
is a matrix Green's function in
both the spinor space  and the spin space.  The ``mass''
$m(k)$ is related to
the sublattice magnetization $M$ by
\begin{equation}
M\sim  \int \frac{d^3q}{(2\pi)^3} {\rm tr}_{\gamma,\tau}[{\bf G}(q)\tau_3].
\end{equation}
Self-consistent equation for $m(k)$ can be obtained from the
 Schwinger-Dyson equation.
Expressed in terms of matrix (both in $\tau$ and $\gamma$ space) Green's
function and (matrix) self energy (pictorially represented by Fig.\ref{sd}a),
the S-D equation is
\begin{equation}
{\bf \Sigma}(k) = -m(k)\tau^3=
-\int \frac{d^3q}{(2\pi)^3}\,\tilde{\Gamma}_{\mu}(k,q){\bf  G}(k-q)
\gamma_{\nu}
\tilde{D}_{\mu\nu}(q)
\end{equation}
where $\tilde{\Gamma}_{\mu},\tilde{D}_{\mu\nu}$
 are fully renormalized vertex and gauge propagator.

  Within the  approximation of replacing
$\tilde{\Gamma}_{\mu}$ by $\gamma_{\mu}$ and $\tilde{D}_{\mu\nu}$  by
$D_{\mu\nu}$, we have
\begin{equation}
m(p)=
\int\frac{d^3k}{(2\pi)^3}\frac{\gamma_{\mu}m(k)\gamma_{\nu}}{k^2+m^2(k)}
D_{\mu\nu}(p-k).
\label{selfc}
\end{equation}
  Still this is a nonperturbative theory; it is easy to check that a finite
order perturbation theory cannot generate a mass term in the fermion
Green's function.  Diagrams that
contribute are shown in Fig.\ref{sd}b.

  The context in which the self-consistent
equation arises is similar to the SDW (spin density wave) problem and
the superconductivity\cite{nambu1}.  However, in our case, the mass $m(p)$
is dependent upon 3-momentum.
Eq.\ref{selfc}
 has been already  analyzed by Appelquist and coworkers in a different
context (``chiral symmetry breaking'')\cite{appelquist}.
Some of the steps are sketched below.

From  the result
\begin{equation}
\gamma_{\mu}\gamma_{\nu}\left(\delta_{\mu\nu}-\frac{q_{\mu}q_{\nu}}{q^2}
\right)=\gamma_{\mu}\gamma_{\mu}-\frac{q\gamma q\gamma}{q^2}=3-1=2
\end{equation}
we have, after some angular integrals,
\begin{equation}
m(p)=\frac{4}{N\pi^2 p}\int_0^{\Lambda}dk \frac{k m(k)}{k^2+m^2(k)}
(k+p-|k-p|).
\label{selfc2}
\end{equation}
Note that the lattice origin of our theory sets the UV cutoff scale $\Lambda$,
while in the theory of Appelquist {\it et al.}
which retains the kinetic term $\sim\frac{1}{4e^2}F_{\mu\nu}^2$,
 the coupling constant $e^2$ sets the
scale (QED3 is a superrenormalizable theory).
The integral equation (Eq.\ref{selfc2})
is equivalent to the differential equation
\begin{equation}
\frac{d}{dp}\left(p^2\frac{dm(p)}{dp}\right)=-\frac{8}{\pi^2N}\frac{p^2m(p)}
{p^2+m^2(p)}
\end{equation}
with boundary conditions
\begin{equation}
\Lambda m'(\Lambda)+m(\Lambda)=0
\end{equation}
and
\begin{equation}
0\leq m(0) < \infty.
\end{equation}
It turns out that this nonlinear differential
equation has a nontrivial solution
only for $N < N_c= 32/\pi^2$\cite{critical}.
 For the physical case of SU(2) antiferromagnet,
$N$ equals 2; therefore,
the dynamical mass generation occurs, and the N\'{e}el-vector rotation symmetry
is spontaneously broken.  Thus, provided that we include the gauge
fluctuations, we have a N\'{e}el order.

The foregoing
argument, however, should be taken with a grain of salt.  In principle,
the gauge propagator that enters the Schwinger-Dyson equation must be a fully
dressed one, and so should be the vertex.  In the symmetry-broken phase,
the gauge propagator is different from the symmetric phase, as the
polarization function $\Pi_{\mu\nu}$ is different.
At the crudest level, if we assume that the fermions acquire a constant
mass $m(p)=m$, then the polarization function would be (See Appendix A).
\begin{equation}
\Pi_{\mu\nu}(q)\sim \frac{1}{m} q^2\left(\delta_{\mu\nu}-
\frac{q_{\mu}q_{\nu}}{q^2}\right),
\end{equation}
which means that a kinetic term $\sim\frac{1}{m}F_{\mu\nu}^2$ is
generated.  This results in the Coulomb potential for the fermions which
in  2+1D is
\begin{equation}
V({\bf x})= -m\int d^2{\bf q}
 \frac{e^{i{\bf q}\cdot {\bf x}}}{{\bf q}^2}
         =  m \ln\left(\frac{|{\bf x}|}{R}\right), \,\,\,\, (R\sim 1/m).
\end{equation}
Since the potential increases at large distances, it is a confining
potential (Sec. V.A),
and a different physical picture for the fermions might be
expected in that case.

Arguments can be made that these considerations do not affect
seriously  the conclusion
as far as the issue of determining {\it whether} dynamical mass generation
occurs and (in the case of occurrence) the value of $N_c$ \cite{maris}.
(The point is that very near $N_c$, the polarization of the fermions in
the symmetry-broken state must be pretty close to that of the massless
fermions. On the other hand, above treatment is too crude to
study the behaviors of  quantities like $m(p)$.
Lattice gauge theory
 simulation\cite{dagotto} does find that the symmetry breaking occurs,
and $N_c\approx 3.5$, which is close to above analytical results.

\begin{figure}
\epsfxsize=\columnwidth
\epsfbox{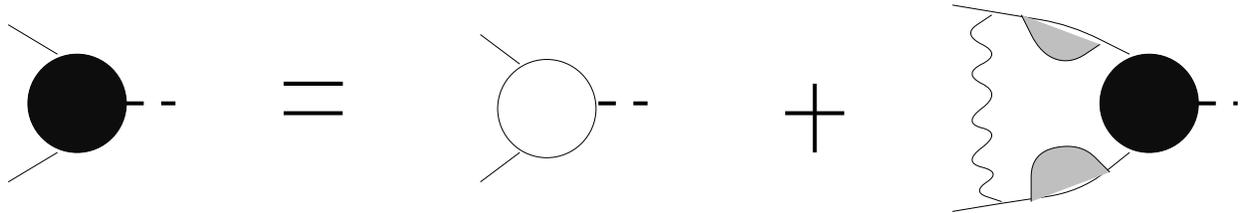}
\caption{Bethe-Salpeter equation for the isovector vertex in the
${\bf Q}=(\pi,\pi)$ channel.  The fermion lines willed a shaded blob
represent the renormalized (self-consistent) Green's function.}
\label{bs1}
\end{figure}

Having seen the symmetry breaking, we can identify the elementary
excitations --- the Goldstone bosons.  We all know that the Goldstone bosons
in an ordered antiferromagnet are spin waves.
In the fermion picture, the Goldstone bosons are a collective mode, and
appear  as a pole in the
two-particle Green's function,
the scattering amplitude  in the appropriate channel,
 and in the related
vertex\cite{vaks}.  Here we show this by considering the SU(2)
vertex.  The Bethe-Salpeter equation for the vertex is given by
\begin{equation}
\Lambda^l(p;q)= 1\cdot\tau^l
- \int\!\! \frac{d^3k}{(2\pi)^3}
 \gamma_{\mu}{\bf G}(k)
\Lambda^l(k;q)
{\bf G}(k+q)\gamma_{\nu}D_{\mu\nu}(p-k),
\label{bsinhomo}
\end{equation}
which is represented by the ladder diagrams of Fig.\ref{bs1}
 (In Eq.\ref{bsinhomo},
the 1
 in the first term on the RHS is the unit matrix in the spinor space.)

If there is a Goldstone boson, then $\Lambda^l(p;q)$ has a pole
at $q^2=0$, in which case  the  homogeneous equation
\begin{equation}
\Lambda^l(p;0)= -\int \frac{d^3k}{(2\pi)^3} \gamma_{\mu}{\bf G}(k)
\Lambda^l(k;0)
{\bf G}(k)\gamma_{\nu}D_{\mu\nu}(p-k)
\end{equation}
has  a nontrivial solution. It is easy to see that
\begin{equation}
\Lambda^l(p;0)=[m(p)\tau^3, \tau^l]
\label{goldstone}
\end{equation}
with $m(p)$ given by Eq.\ref{selfc}
is the solution.  Therefore, if there is a dynamical mass generation
($m(p)\neq 0$), then we have two Goldstone bosons (SU(2) symmetry
breaking): $\tau^l=\tau^1,\tau^2$ (or $\tau^+,\tau^-$)
 in Eq.\ref{goldstone} gives a
 nonvanishing commutator.  The Lorentz invariance
restricts the mesons to have a linear dispersion  $q_0^2={\bf q}^2$
(Minkowski space),
which is indeed the case with antiferromagnetic spin waves.

\section{Two-dimensional Underdoped Cuprates}
\subsection{antiferromagnetic correlations}
The foregoing may appear to be an unnecessarily roundabout
 way  of looking at 2d quantum antiferromagnet,
yet we believe it is not without certain value.  The qualitative idea that
a strong attraction between spinons and antispinons via gauge field can result
in the formation of a vector condensate with {\bf Q}=($\pi,\pi$)
(the antiferromagnetic
channel)  may shed some lights on
the underdoped cuprates.  In the underdoped cuprates, an effective theory
based on the sFlux ansatz of the SU(2) mean
field theory\cite{kimleewen,wenlee,leeetal}
consists of 2 flavors of massless Dirac fermions and a U(1) gauge field
just like above, but now the gauge field is also coupled to the bosons
(holons).  In other words, schematically,
\begin{equation}
L= \bar{\psi}_{\alpha}\gamma_{\mu}
(\partial_{\mu}
-ia_{\mu})\psi_{\alpha} - ia_{\mu}J_{\mu}^B + L_B.
\end{equation}
 This additional coupling to the bosons will weaken the gauge
field in the sense that it will screen the time component of the gauge field
($\lim_{q\rightarrow 0}\Pi_{00}^B\neq 0$).
The gauge propagator then would take  the form (in the Coulomb gauge)
\begin{equation}
D_{\mu\nu}\sim \frac{1}{\sqrt{q^2}}\delta_{\mu i}\delta_{\nu j}
\left(\delta_{ij}-\frac{q_iq_j}{{\bf q}^2}\right)
+\frac{1}{x\tilde{J}+\sqrt{q^2}}\delta_{\mu 0}\delta_{\nu
0}\frac{q^2}{{\bf q}^2}
\label{lnon}
\end{equation}
where $i,j=1,2$ are spatial indices and $x$ is the concentration of doped
holes.
In the simplest approximation, we ignore the massive part, and consider
only the spatial component
\begin{equation}
D_{\mu\nu}(q)=\frac{8}{N\sqrt{q^2}}
\delta_{\mu i}\delta_{\mu j}\left(\delta_{ij}-
\frac{q_iq_j}{{\bf q}^2} \right),
\end{equation}
and examine the self-consistent equation for $m(p)$ (Eq.\ref{selfc}).

Since
\begin{equation}
\gamma_{\mu}\gamma_{\nu}D_{\mu\nu}=\gamma_i\gamma_i - \frac{(q_i\gamma_i)
(q_i\gamma_i)}{{\bf q}^2}= 2-1 =1,
\end{equation}
 we have
\begin{equation}
m(p)=\frac{2}{N\pi^2 p}\int_0^{\Lambda}dk \frac{k m(k)}{k^2+m^2(k)}
(k+p-|k-p|).
\end{equation}
This is identical to the Eq.\ref{selfc}, except for a factor of 2 difference
in the prefactor.  This factor 2 reduction can be understood from that fact
that the gauge
field in 2d has one transverse mode and one longitudinal mode the latter of
which becomes massive.  From the analysis following
Eq.\ref{selfc}, we know immediately that there
will be a symmetry breaking only for $N < N_c'=N_c/2=16/\pi^2$.
Thus, for the physical case of $N=2$,
 the spontaneous symmetry breaking would not occur!---This is what was
hoped for our mean field theory.

\begin{figure}
\epsfxsize=\columnwidth
\epsfbox{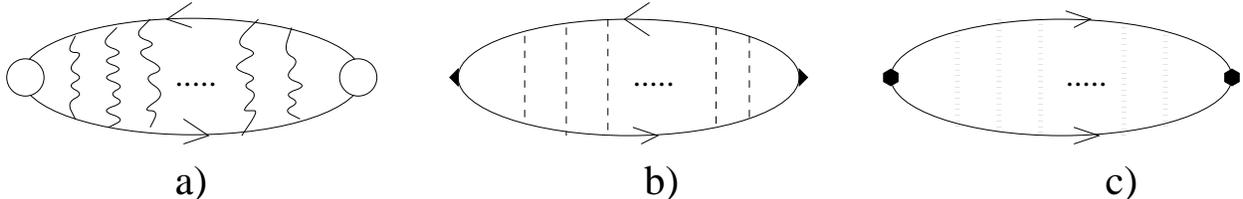}
\caption{Ladder diagrams. a) Our case: staggered spin correlation.  The
wiggly lines are interactions mediated by the gauge field.  Structurally
similar examples:  b) ferromagnetic
spin correlation.  The dashed lines are short-range repulsive interactions.
c) superconducting correlation.  The dotted lines are some kind of attractive
interaction causing pairing.}
\label{ladders}
\end{figure}

The attraction in the ${\bf Q}=(\pi,\pi)$ channel
mediated by the
gauge field, although not strong enough to generate a condensate,
 will nevertheless have
a strong effect on the
spectrum of antiferromagnetic excitation.   The fluctuation of the
order parameter associated with the transition (staggered moment)
can be examined by looking at the staggered-channel
spin correlation function in the ladder approximation,
similar to the more familiar
problems like  the
superconducting fluctuations or the ferromagnetic spin
fluctuations\cite{hertz} (See Fig.\ref{ladders}).

  In the problem of nearly ferromagnetic Fermi liquids,
short range interaction between fermions
are often modelled in terms of an on-site repulsion  $U$
($UN(E_F)$ is the dimensionless coupling constant corresponding to our $1/N$).
This problem is a lot simpler, as the ladder series sums immediately
to the RPA form $\chi=\chi_0/(1-U\chi_0)$.  The pole of the RPA propagator
gives the diffusive mode (``paramagnons'')
associated with a conserved order parameter.  This mode (more accurately
the peak in $\chi''(\omega)$)  comes down in energy and becomes sharper
 as  $U$ approaches $U_c$.

  Unfortunately, in our problem
the interaction is retarded and long-ranged,
 hence  the diagrams are not easy to evaluate.
   Nevertheless, on physical grounds,
 it is quite reasonable to expect that the same gauge field
which  caused the antiferromagnetic instability in the absence of holons
will try to create a (massive) mode
(particle-hole bound state in the ${\bf Q}=(\pi,\pi)$
channel) in this case.  Because the symmetry is
unbroken, there is  a particle-hole
continuum, as a result of which
a sharp mode cannot exist, but a ``broad resonance'', {\it
i.e.} a very unstable
meson with a Minkowski-space  dispersion
\begin{equation}
q^2=\tilde{m}^2-i\tilde{m}\Gamma,
\end{equation}
is expected.
 The mass of the mode $\tilde{m}$   would come down
as the transition is approached (``soft mode'').
 A physical consequence will be that
the dynamic  susceptibility
$\chi_{{\bf Q}=(\pi,\pi)}''(\omega)$ (which can
be probed by neutron scattering)  has
a broad peak, with a substantial rearrangement of the spectral weight
compared to the mean field prediction.  This  heuristic picture
is consistent with experiments of Keimer and collaborators\cite{keimer}
that find in the normal state (and in the superconducting
state) of  underdoped cuprates  a magnetic scattering with a broad
peak at some frequency scale that comes down in energy as the doping
is reduced.

\begin{figure}
\epsfxsize=\columnwidth
\epsfbox{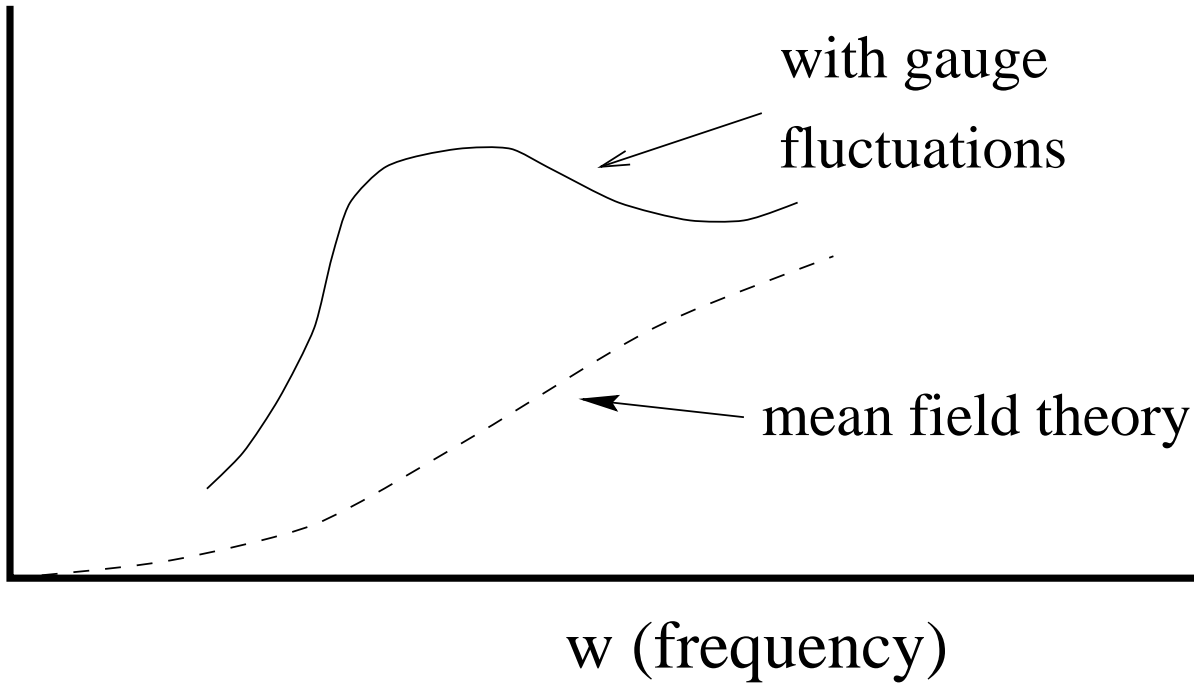}
\caption{Rearrangement of spectral weight in 
$\chi_{{\bf Q}=(\pi,\pi)}''(\omega)$ due to gauge fluctuations.}
\label{chi2}
\end{figure}

 In principle, we should be able to study this by looking into the
 Bethe-Salpeter equation for spinon-antispinon scattering amplitude or
 the associated vertex.  The analysis, however,
 entails a number of practical and
technical difficulties.  We can get around the problem of the complicated
Lorentz-noninvariant gauge  propagator for the doped case
(Eq.\ref{lnon}) by convincing ourselves that the we would obtain a similar
physics by considering  a Lorentz invariant one
\begin{equation}
D_{\mu\nu}= \frac{8}{\bar{N}\sqrt{q^2}}\left(\delta_{\mu\nu}-\frac{q_{\mu}
q_{\nu}}{q^2}\right)
\end{equation}
with $\bar{N}> N_c$ (Reducing the doping would correspond to reducing
$\bar{N}$).
Still, the fact that we are investigating a mode that
is {\it massive and damped}  causes complications unseen in the B-S equation
for the Goldstone bosons of the ordered antiferromagnet.
  Generally, the scattering amplitude  and the
vertex have many components (``invariant amplitudes''), {\it e.g.},
\begin{equation}
\Lambda^l(p;q)= f(p;q)\tau^l+ g(p;q)p\gamma\tau^l+
\cdots .
\end{equation}
The Goldstone modes are massless, hence we can focus on $q=0$ in which case
$f(p;0)$ decouples from other amplitudes and has the same  equation
as that of the dynamical mass, as we have seen in Sec.III.C.  The
decoupling doesn't occur for $q\neq 0$.
  Appelquist
{\it et al.}\cite{appel_nv}  have
considered the scalar component of the Euclidean scattering amplitude in the
ladder approximation, ignoring the coupling to other components;
 this corresponds to considering $f(p;q)$ only.
They claim to find no light meson  whose mass
comes down to zero as the transition is approached, and hence the
transition must be a novel one, {\it i.e.} neither a first order
nor a second order
transition.  However, this conclusion is questionable, as
a well-defined mode (pole in the scattering amplitude)
 in the symmetric phase is too stringent a requirement
for a second order phase transition; more generally what happens in
the symmetry-unbroken phase as the second order
 transition is approached is a build-up of the
low energy spectral weight for the  response related to the order parameter.
In any case, the analysis of Ref.\onlinecite{appel_nv} is not appropriate for
finding an unstable meson which resides in the second Riemann sheet of
the Minkowski space.

We now discuss briefly the NMR relaxation rates, which measure the  very low
energy limit of the spin excitation spectrum.  As noted earlier, the mean
field theories predict similar behavior of the Oxygen and Copper site
relaxation rates, while experiments show marked difference.
It would be  difficult to
calculate  finite temperature properties within
our gauge theory, but we believe that what has been discussed so far throws
some light on this issue: once the gauge fluctuation is included,  there
is no reason why the two relaxation rates should have similar behaviors,
since the effect of the gauge field on
the {\bf Q}=0 and {\bf Q}=$(\pi,\pi)$ response is very different.
It is plausible that at high temperature side the Cu
$1/T_1T$
will be increasing with decreasing temperature, as the gauge fluctuation
restores certain amount of antiferromagnetic correlations (note
that above some energy scale, the screening effect of bosons won't be too
important and the propagator would look like that of the undoped cuprates),
while at low
temperatures (below some scale) it will go down with decreasing temperature,
since we still have the d-wave gap if the mean field picture is going to
survive.

\begin{figure}
\epsfxsize=\columnwidth
\epsfbox{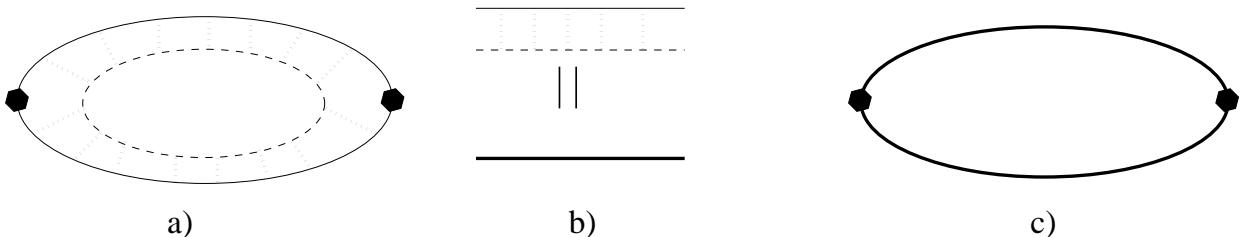}
\caption{Possible contribution from boson-fermion coupling. a) The inner loop
(dashed line) and the outer loop (solid line) denote  bosons and fermions,
respectively, and the dotted lines denote boson-fermion interaction
b) ``electron'' propagator c) schematic representation of the same diagram}
\label{bosfermi}
\end{figure}

Finally, we note that in our discussion of antiferromagnetic excitations in
underdoped cuprates we have not considered
 certain class of diagrams that may be important. An example, shown
in Fig.\ref{bosfermi}, involves a boson loop and  a fermion loop exchanging
a residual attractive interaction.  The set of a boson
and a fermion line connected by ladder type interaction can be viewed as
a electronlike propagator (Bethe-Salpeter equation for electron propagator).
This type of contribution may be related to
 pieces of the Fermi surface near $(\pm \pi/2,\pm \pi/2)$\cite{wenlee},
and might provide some clue
to the origin of low-frequency
incommensurate features observed in the neutron scattering.

\subsection{thermodynamic properties}
In the underdoped cuprates the gauge field interaction also  affects
the uniform part of the spin response (the thermodynamic properties)
though the effects are subtler.  More specifically, the  coupling
of the gauge field to nonrelativistic
 bosons results in the logarithmic renormalization of the
velocity of the fermions, which has the effect of enhancing the specific heat
and the uniform susceptibility\cite{kimleewen}.
  This seems to be in accordance with
experiments on underdoped cuprates.  Although one might alternatively
view the enhancement features in the normal state as having to do
with the ``Fermi surface segments'', these are not really quasiparticle
states in the strict sense ($z$=0) and there are likely to be theoretical
complexity and possibility of overcounting.  It seems a lot simpler to view
that the fermions account for most of the entropy and spin response.
 A curious feature\cite{sachdev_pr} of our theory
is that the Wilson ratio $W=C/T\chi_u$
 has a value quite close to
that of the ``quantum critical'' phase  of the nonlinear sigma
model\cite{chubsach}: $W$ in our theory is 0.128, while the
$O(N)$ nonlinear sigma model gives $W=0.124$ at zeroth order, and
$W=0.116$ with the inclusion of $1/N$ correction.  Could this be
a coincidence?

\section{Concluding Remarks}
We now conclude with a discussion of
 some difficult matters, hoping not so much as to resolve them
but to content ourselves with
placing them in perspective.

\subsection{confinement, spinons, and all that}

The term ``spinon'' has been used rather carelessly in this paper.
In the strict sense, it refers to well-defined neutral elementary
excitations carrying spin 1/2.  A well-known (and perhaps the only
widely-accepted) example is the
solitonic (topological)
objects in spin chains that are usually understood within the bosonization
framework\cite{faddeev,tsvelik}.

The ``spinons'' in this paper refer most of the times
to spin 1/2 fermions strongly
coupled to fluctuating gauge fields --- the fermions that arise
from the RVB mean field theories.  These spinons are far from being
well-established.  In 1d, the fact that they are different from
 topological spinons (``true spinons'') has been emphasized
by Mudry and Fradkin\cite{mudry2}.
 Nevertheless, we have seen that in the 1d Heisenberg model, the mean field
fermions give a reasonable description of the specific heat and uniform
susceptibility\cite{sebas};
the Zeeman splitting of the spin up and down fermions in a
magnetic field and the
associated incommensurate magnetic excitations\cite{reich} can be easily
understood in terms of the
mean field fermions.  Furthermore, we showed in Section II that
the perturbative treatment of the
gauge fluctuations systematically improves the antiferromagnetic
correlation.  Thus the mean field fermions are not as bad a
starting point as some would regard.

In 2d and 3d, pre-high \Tc empiricism and experiences
  provide  strong resistance to the notion of spinons as derived from RVB
  mean field theory.  After all,
in most metals (2 and 3d), the basic elementary
excitation is  quasiparticles that carry both spin 1/2 and charge $e$, and
in most insulating antiferromagnets or in spin liquids (like spin ladders),
the elementary
excitations are spin 1 objects (spin waves or magnons).   In other words,
spinons  and holons do not exist on their own,
 but are ``confined'' in  spinon-antispinon
composite objects (spin waves: $f^*f$) or  in  spinon-holon composite objects
(quasiparticles: $f^*b$).

The sense in which this ``confinement'' is discussed is  similar
to the  quark confinement in particle
physics ---
the absence outside the nucleus of the
 quarks that make up hadrons (baryons
and mesons).  In fact,  the strong interaction physics  appears to
share quite a few parallels with our problem\cite{laughlin_cf}.
Quantum chromodynamics (QCD),
which is widely believed to be the correct theory of
 strong interactions, is a gauge theory
whose low energy physics is  as poorly understood
as the high \Tc cuprates.  At a more substantial
level, the phenomenology of strong interactions indicates
 that the (approximate)
 chiral symmetry is spontaneously broken, giving rise to mesons (such
as pions) which are Goldstone bosons, in analogy with spin waves in
a N\'{e}el ordered system.  Just as the
nonlinear sigma model
gives an excellent description of the low energy spin dynamics of the undoped
cuprates\cite{CHN},
 in particle physics it has been well known that similar effective
lagrangians (sigma models, chiral perturbation theory\cite{gasser},
 etc.)
 give a very good description
of the hadronic physics.   However, attempts to
``derive'' the parameters
of the effective theories, such as the pion decay constant, from first
principles
have not been entirely successful.  The basic difficulty is
that the same asymptotic freedom that led to the confirmation of the
quark picture at the high energy side causes grave difficuties in analyzing
the low
energy physics.   At present, no consensus exists as regards the ``mechanism''
 of
chiral symmetry breaking in QCD,
 but at least it seems clear that the problem is
intimately connected to the general issue of confinement.

It is the basic  underlying idea  of this
paper  that the confinement does
not occur (at least)
for the normal state of the superconducting cuprates, or more
loosely, spinons and holons are ``meaningful'' objects.
 On the other hand,
{\it some sort of} confinement is relevant to the discussion of
the N\'{e}el ordered state in undoped
cuprates, since  we
know that there the low energy excitations are not the spinons but
spin waves.  In considering the confinement, we are (helped but also)
 burdened by previous studies in particle physics
regarding the issue.  The confinement motivated by the strong interaction
phenomenology --- no color nonsinglet particles exist in the
physical spectrum --- is a  strong statement,
 and it is not
clear to what extent the confinement in our case should match
 the quark confinement.

For example, we can ask the following questions in 2d undoped systems.
1) On the low energy side,
are the spin waves the only massless degrees of freedom, or could there be
an additional mode hiding?  2)  On the high energy side, do the spinon states
exist, or have they disappeared completely?  In other words, can we somehow
``see'' spinons at high energies?

To address these questions, we need to discuss  one possibly
important aspect of our gauge theory, the compactness, that has been ignored
so far.  The models that we have examined in this paper
originate from the lattice, hence are compact gauge theories.
The usual assumption is that the compact theory can be replaced by a more
amenable  noncompact theory in the continuum, but this is not always justified.
A representative and scary example is the pure gauge theory in 2+1D.
Polyakov\cite{polyakov}
has shown that the compact pure
gauge theory (2+1D electrodynamics)
differs from the noncompact one due to instantons ---
 the  topologically nontrivial, extended
 classical solutions of Euclidean
gauge field equations, that can be viewed as tunneling events between
topologically inequivalent vacua.  Instantons cause
 the Wilson loop
 to follow the area law $\langle\exp(\oint dx_{\mu}
a_{\mu})\rangle\sim \exp(-Area)$,
which means the presence of a linear potential
$V({\bf x}^a,{\bf x}^b)\sim |{\bf x}^a-{\bf x}^b|$ between static sources,
a sign of  (strong) confinement.

When matter fields are present as in our case and in QCD,
the situation is not so simple.  In QCD, for example, despite theoretical
attempts\cite{CDG}  the relevance of
instantons to quark confinement remains uncertain.  Generally, the
fluctuations of matter fields (especially the massless fields) are adverse
to instantons.  One specific scenario by which instantons are suppressed is
fermion zero modes.  For example, in the massless Schwinger model, the
fermion determinant ${\rm det}[(\partial_{\mu}-ia_{\mu})\gamma_{\mu}]$
in a gauge configuration $a_{\mu}$ with an instanton vanishes, so
only the topologically trivial sector contributes to the functional intergral
(partition function).
 Similar mechanism turned out to account for
the famous U(1) problem\cite{thooft} in QCD.
In both cases (1+1D and 3+1D), the zero mode is  connected
 to (axial) anomalies, whose analogue in odd space-time dimensions
(2+1D) is nonexistent.

In the context of our problem (compact 2+1D gauge theory with Dirac fermions),
Marston\cite{marston} studied the possibility of fermion zero modes, and
concluded that zero modes do not exist, suggesting a
 possible relevance of instantons.
Marston has also calculated the action of an instanton, and found that it
is logarithmically divergent with a prefactor proportional to the
number of flavors ($S\propto N\ln R$).  Kosterlitz-Thouless type argument
then indicates that below critical ${\cal N}_c$ which turns out to be 0.9,
instantons
may proliferate, while for  $N > {\cal N}_c$ which includes the physical
case ($N=2$), the instantons are suppressed.
Ioffe \& Larkin\cite{IL} have also calculated
 the action of an instanton, considering only the bilinear part of the
gauge field action obtained by integrating out the fermions,  and found
 logarithmic  divergence, but ${\cal N}_c$ in this case turns out to be 24.
It has been noted that this treatment may be too simple, and the question
of instantons for physical $N$ remains unclear\cite{khlebnikov}.

{\it The suppression of instantons
 may no longer be the case if the fermion masses are dynamically
generated by spontaneous symmetry breaking,
 which is indeed our situation.}  We now have the possibility of
symmetry-breaking-induced instantons.
 These instantons here  cannot induce the dimerization
(valence bond solid order)\cite{marston3}, unlike those of the bosonic spinon
theory\cite{readsach}.  {\it Probably the most notable
consequence of the instantons
is that the gauge field
will become  massive}\cite{polyakov}, {\it hence the spin waves will be the
{\rm only} low energy excitations of the ordered antiferromagnet.}

On the other hand, in the (simpler) scenario without the instantons
(Sec.III.C),
there is a massless gauge field in addition to the spin waves in the
ordered antiferromagnet.
 This is rather bothersome, since it is commonly believed
that the spin waves deplete the low energy excitations in 2d quantum
antiferromagnet.  Practically, it may be difficult to determine the absence
or the existence of the massless  gauge field, as it does not couple
to external probes in a simple way.
The gauge field may have a
$\sim T^2$ contribution to the specific heat like the spin waves, but
the modification of the prefactor due to gauge field might be considered a
spin-wave renormalization effect.

The pictures with or without instantons may also have different implications
for the spinon confinement.  If the instantons are relevant, the Polykov-type
nonperturbative mechanism might lead to the final picture in which the
spinons have vanished completely, and traces of them might not be present
even at high energies.  In the picture without instantons, spinons may
be still confined, due to the long range nature of the gauge field.  A
situation in which this works out\cite{witten} is the 1+1D
$CP^{N-1}$ model ($L=|(\partial_{\mu}-
ia_{\mu})z|^2 +m^2z^{\dagger}z$), in which integrating out the
 matter field ($z$-field) generates a kinetic term (self energy) for
the gauge field $\frac{1}{m} F_{\mu\nu}^2$ that gives rise
to the linearly confining Coulomb potential
({\it i.e.} $V(x_1)=-\int dq_1 \exp(iq_1x_1)/q_1^2
\sim |x_1|$) between the $z$-charges.  In our 2+1D problem (Sec.III.C),
at a crude level, the dynamical mass generation will have a feedback
effect on the gauge field, softening the gauge propagator  from
$1/\sqrt{q^2}$ to $m/q^2$, making it look like a genuine electromagnetic
field.  In 2+1D, the Coulomb potential
associated with electromagnetic field is weakly (marginally)
confining  ($V({\bf x})\sim \log(|{\bf x}|/R)$)\cite{weneffective}, and
for distances \{energies\} shorter \{higher\} than some scale $R\sim 1/m$
\{$m$\},
the spinon might be ``visible''.  Within the framework of Sec.III.C, to see
if there are well-defined spinons, we can study whether the Minkowski-space
 Green's function of the spinons has a pole at some mass scale.
If the dynamical symmetry breaking occurred in such a way that the spinons
acquire a constant mass as in the NJL model\cite{nambu2}, then the spinons
should exist at high energies, and we might be
able to ``see'' them\cite{magassume} by a method analogous to the 1d
example\cite{reich}.
 If the symmetry is broken by (long-range,
retarded) gauge field interactions as in our case,
the problem is quite complicated;  we need to continue
the mass function $m(p)$
from the Euclidean space to the
Minkowski space ($p^2\rightarrow -p^2$), and examine whether $p^2-m(-p^2)=0$
has a solution near the real axis.  The results from the literature\cite{maris}
 indicate that when the ``feedback'' effect of the spinon mass generation on
the vacuum
polarization is taken into account (in which case the gauge field dynamics
looks like $F_{\mu\nu}^2$), no poles are found near the real axis.

At present, it is not clear
 which picture ({\it i.e.} with or without instantons) of
confinement  is realized for the undoped cuprates.
In the normal state of the underdoped cuprates, instantons are probably
not relevant,
as we have massless fermions and there are additional massless fluctuations
due to the introduction of holes.
In any case, the unusual phenomenology of
high \Tc cuprates points to that the spinons and holons are deconfined,
as emphasized earlier.
 It is beyond the scope of this paper to
discuss the issue of possible confinement in the superconducting state.

\subsection{loose ends}
In this paper we have examined the magnetism of undoped and
underdoped cuprates
 from the point of view of
neutral fermions with spin 1/2\cite{yulu}.
Admittedly, the theory as it stands is far from rigorous.
The philosophy has been
to analyze possibly the simplest
 effective field theory of massless Dirac fermions, bosons and
gauge fields,
motivated by the sFlux phase that appears as a saddle point solution
of the SU(2) mean field theory.  In reality, the situation is a lot more
complicated.  The mean field fermion
spectrum is anisotropic: $\epsilon({\bf k})= \sqrt{v_F^2\tilde{k}_1^2
+v_2^2\tilde{k}_2^2}$, and the velocities
$v_F, v_2$ have some doping and temperature dependences.
These dependences, however,
do not account for the puzzling properties of the cuprates that
we have discussed, and
 the photoemission does  indicate that
the quasiparticle gap remains large in the normal state of
the underdoped cuprates ({\it i.e.} superconducting
transition is not a gap-closing transition) and the gap is only weakly
doping dependent.  Therefore, the effective theory may be quite sensible
for studying qualitative features not captured by mean field theories.
We have made a number of approximations to treat the complicated
 dynamics of the
gauge fields, but we hope it's not {\it too}
 optimistic to  view that the qualitative conclusions
are correct.  In any case, the
idea that the holon coupling to the gauge field prevents spontaneous
symmetry breaking in the fermion system is an attractive
one, and is very much
in the spirit of the empirical fact that moving holes quickly destroy
antiferromagetic order.  Unfortunately, even with simplifications, the
calculations quickly become rather intractable.

Eventually the ``spin gap''
 has to close up as we go to the optimally doped regime,
which in the mean field theory is modelled by the uRVB saddle point\cite{PN}.
  The
details of this crossover is certainly beyond our hopes.  This will
severely limit us in considering some very interesting issues, like the
relation between the neutron scattering peaks in the underdoped cuprates
and  the sharp 41meV  peak in the superconducting
phase of the optimally doped ${\rm YBCO_7}$.  Anyway, the present
theory does not say much about the spin excitations {\it in the
superconducting state} of the underdoped cuprates, since the dynamics of
gauge field in the superconducting state is different from that
considered here.
 Another drawback is that
in our framework it is not easy to consider possible
incommensurate features in spin excitations.

Granting these limitations, we still  feel that the picture of
spin excitations in underdoped cuprates in terms
 of deconfined
fermionic spinons is reasonable and perhaps
 more natural than other descriptions
like those based on  fluctuating staggered moments.
Features like the linear-in-$T$ behavior of uniform
susceptiblity and the specific heat coefficient in underdoped cuprates
 might  be also explanable in terms of the
 ``quantum critical'' regime of the
nonlinear sigma model,
but it is not clear in that approach
how to account for the strange behavior of the Copper $1/T_1T$
in the same temperature range.
 Attempts to explain the Copper $1/T_1T$ in terms of the
``quantum disordered'' regime then has to explain why activated behaviors
are not seen in quantities like  uniform susceptibility. Again, this
seems to point to the difficulty of achieving a theoretical description
of a system that involves a mysterious combination of  gaplike (short range)
 correlations and  critical correlations.

\section{acknowledgments}
We would like to thank X.-G. Wen for collaborating on an earlier project that
led to this work, and for many helpful conversations.
Helpful discussions or correspondences  with
A. Furusaki,
F. D. M. Haldane, D. A. Ivanov, C. Mudry,  N. Nagaosa, K. Rajagopal,
 S. Sachdev, and  O. Syljuasen, are also  gratefully
acknowledged.  This work was supported by the MRSEC Program of the NSF
under award number DMR
94--00334.

\begin{appendix}
\section{Fermion Polarization}
In this appendix, fermion vacuum-polarizations in 1+1D and 2+1D are
worked out using
``dimensional regularization''.

For massless Dirac fermions, the
 polarization function $\Pi_{\mu\nu}$ is given by
\begin{eqnarray}
\Pi_{\mu\nu}(q)&=&-N\int \frac{d^Dk}{(2\pi)^D}\,
{\rm tr}[G(k)\gamma_{\mu}G(k+q)\gamma_{\nu}]
\nonumber \\
&=&N\int \frac{d^Dk}{(2\pi)^D}\,
 {\rm tr}\left[\frac{k\gamma}{k^2}\gamma_{\mu}\frac{(k+q)\gamma}
{(k+q)^2}\gamma_{\nu}\right] \nonumber \\
&=&N\int_0^1 dx \frac{d^Dk'}{(2\pi)^D}
 \frac{{\rm tr}[(k'+(1-x)q)\gamma \gamma_{\mu}
(k'-xq)\gamma \gamma_{\nu}]}{(k'^2+q^2x(1-x))^2},
\end{eqnarray}
where $N$ is the number of flavors of fermions.
Using the trace identity
\begin{equation}
{\rm tr}[\gamma_{\rho}\gamma_{\mu}\gamma_{\sigma}\gamma_{\nu}]
=({\rm tr 1})(\delta_{\rho\mu}\delta_{\sigma\nu}-
\delta_{\rho\sigma}\delta_{\mu\nu}
+\delta_{\rho\nu}\delta_{\mu\sigma}),
\end{equation}
The numerator of the integrand is found as
\begin{equation}
2k_{\mu}'k_{\nu}' -2x(1-x)(q_{\mu}q_{\nu}-\delta_{\mu\nu}q^2)
-\delta_{\mu\nu}[k'^2+q^2x(1-x)].
\end{equation}
Substituting, we get
\begin{eqnarray}
\Pi_{\mu\nu}(q)=N ({\rm tr 1})
 \int_0^1 dx\int \frac{d^Dk}{(2\pi)^D}\left[\frac{2k_{\mu}k_{\nu}}{(k^2+
q^2x(1-x))^2}  \right. \nonumber\\
\!\!\!\!\!\left.\!\!\!\!\!\! -\frac{\delta_{\mu\nu}}{k^2+q^2x(1-x)}
+ \frac{2x(1-x)(\delta_{\mu\nu}q^2-q_{\mu}q_{\nu})}{(k^2+q^2x(1-x))^2}
\right].
\end{eqnarray}
The first two terms cancel, and we have
\begin{eqnarray}
\Pi_{\mu\nu}(q)&=& 2 N ({\rm tr 1})\frac{\Gamma(2-D/2)}{(4\pi)^{D/2}}
(q^2\delta_{\mu\nu}-q_{\mu}q_{\nu})
\int_0^1dx\,
 x(1-x)(q^2 x(1-x))^{D/2-2} \nonumber \\
&=& \left\{ \begin{array}{c}
\frac{N({\rm tr 1})}{32}\sqrt{q^2}\left(\delta_{\mu\nu}-
\frac{q_{\mu}q_{\nu}}{q^2}\right)\,\,\,\,\,\, D=2+1  \\
 \frac{N({\rm tr 1})}{2\pi}\left(\delta_{\mu\nu}-
\frac{q_{\mu}q_{\nu}}{q^2}\right)\,\,\,\,\,\, D=1+1. \end{array}\right.
\end{eqnarray}
For the situations discussed in the main text, we have  ${\rm tr 1}$=4
for the 2+1D (Secs. III \& IV) as we have 4 component fermions, while
${\rm tr 1}$=2 for the 1+1D (Sec. II).

If the fermions are  massive,
{\it i.e.},
$L= \bar{\psi}(\partial_{\mu}-ia_{\mu})\gamma_{\mu}\psi + m\bar{\psi}\psi$,
similar calculation shows that the polarization function is given by
\begin{eqnarray}
\Pi_{\mu\nu}(q)&=&
2 N ({\rm tr 1})\frac{\Gamma(2-D/2)}{(4\pi)^{D/2}}
(q^2\delta_{\mu\nu}-q_{\mu}q_{\nu})\int_0^1dx (1-x)x(m^2+q^2
x(1-x))^{D/2-2} \nonumber \\
&=& \frac{({\rm tr 1})N}{2\pi}(q^2\delta_{\mu\nu}-q_{\mu}q_{\nu})\left(
 \frac{1}{q^2}- \frac{4m^2}{q^3\sqrt{4m^2+q^2}}
\tanh^{-1}(\frac{q}
{4m^2+q^2})\right) \,\,\,\, [D=1+1], \nonumber \\
&=&   \frac{({\rm tr 1})N}{4\pi}     (q^2\delta_{\mu\nu}-q_{\mu}q_{\nu})
\left(\frac{m}{2q^2}+ \frac{q^2-4m^2}{4q^3}
\sin^{-1}(\frac{q}
{\sqrt{4m^2+q^2}})\right) \,\,\,\, [D=2+1].   \nonumber
\end{eqnarray}
For small $q$ (i.e. $q^2 << m^2)$), we would have
\begin{eqnarray}
D=2+1: \Pi_{\mu\nu}(q)&\approx &   \frac{({\rm tr 1})N}{4\pi}
(q^2\delta_{\mu\nu}-q_{\mu}q_{\nu})
 \left(\frac{1}{6m} - \frac{q^2}{60m^3}+..\right)
\nonumber \\
D=1+1: \Pi_{\mu\nu}(q)&\approx & \frac{({\rm tr 1})N}{2\pi}
(q^2\delta_{\mu\nu}-q_{\mu}q_{\nu})
 \left(\frac{1}{6m^2} - \frac{q^2}{30m^4}+..\right). \nonumber
\end{eqnarray}
Note that the  massive fermions give rise to a gauge field dynamics of the
form $L\sim a\Pi a\sim (F_{\mu\nu})^2$ like the real electromagnetic field.

\section{Jacobian for chiral rotation}
In this appendix, we derive  the jacobian $\exp(J)
=\frac{D\bar{\psi}D\psi}{D\bar{\psi}'D\psi'}$ for chiral
rotation, along the line of Stone\cite{stone}.  Basically we require
the  functional integral  $Z=\int D\bar{\psi}D\psi \exp(-L)$
be invariant under the chiral
transform. For an
infinitesimal chiral transform $\psi\rightarrow \psi'=
\exp(-\gamma_5\theta_a)\psi$, the measure changes as
\begin{equation}
D\bar{\psi}(x)D\psi(x)\rightarrow\! D\bar{\psi}(x)D\psi(x) \left(1\!
-\!\int d^2x \frac{\delta J(x)}{\delta\theta_a(x)}\theta_a(x)\right),
\end{equation}
while the lagrangian changes as
\begin{equation}
L(\psi)\rightarrow L(\psi) -(\partial_{\mu}\theta_a)\bar{\psi}
\gamma_{\mu}\gamma_5
\psi= L(\psi) +i\theta_a\partial_{\mu}j_{5\mu}.
\end{equation}
Setting the functional derivative
\begin{equation}
\frac{\delta Z}{\delta\theta_a}
=\left\langle \int d^2x \left(-\frac{\delta J}{\delta\theta_a}
-i\partial_{\mu}j_{5\mu}\right)\theta_a\right\rangle
\end{equation}
to  zero gives
\begin{equation}
\frac{{\delta}J}{\delta\theta_a}=
-\frac{N}{2\pi}\epsilon_{\mu\nu}F_{\mu\nu}
= + \frac{N}{\pi}\partial_{\mu}^2\theta_a,
\label{final}
\end{equation}
where we have used the axial anomaly condition
$\partial_{\mu}j_{5\mu}= -\frac{iN}{2\pi}\epsilon_{\mu\nu}F_{\mu\nu}$
with $F_{\mu\nu}=\partial_{\mu}a_{\nu}-\partial_{\nu}a_{\mu}$,
$a_{\mu}=\epsilon_{\mu\nu}\partial_{\nu}\theta_a + \partial_{\mu}\theta_b$.
 Eq.\ref{final} then implies
\begin{equation}
J=-\int d^2x \frac{N}{2\pi} (\partial_{\mu}\theta_a)^2.
\end{equation}
\end{appendix}


\begin{references}
\bibitem{literature} Such a scenario has been already mentioned in the
literature. See, for example, Refs.\onlinecite{laughzou,marston}.
\bibitem{critical} Note that this number ($N_c$) was derived using the
Landau gauge. $N_c$ (physics) should not depend on the gauge choice,
though approximate calculations using other gauge may give somewhat different
value of $N_c$.
\bibitem{magassume} We will need a big magnetic field comparable
to $J$; It is  assumed
 that high magnetic field simply causes rigid shifts of spinon bands. This
possibility is currently under investigation.
\end{references}
\end{document}